\def\beq{\begin{equation}}
\def\eeq{\end{equation}}
\def\beqa{\begin{eqnarray}}
\def\eeqa{\end{eqnarray}}
\def\eqn(#1){\label{#1}}
\def\eq(#1){\label{#1}}
\def\ceq#1{(\ref{#1})}

\def\F{\Phi}
\def\Ffid{\Phi_{\rm fid}}
\def\drxn{{\bf n}}
\def\lum{\Lambda}
\def\rvec{{\bf r}}
\def\like{{\cal L}}
\def\params{{\cal P}}
\def\sparams{{\cal S}}

\documentstyle[11pt,aaspp4]{article}

\slugcomment{Submitted to {\it The Astrophysical Journal}, December 1996}

\lefthead{Loredo and Wasserman}
\righthead{Gamma Ray Bursts: Anisotropic Models}

\begin{document}

\title{Inferring the Spatial and Energy Distribution\\
of Gamma Ray Burst Sources. III. Anisotropic Models}

\author{Thomas J. Loredo and Ira M. Wasserman}
\affil{Center for Radiophysics and Space Research,
Cornell University, Ithaca, NY 14853-6801}

\begin{abstract}
We use Bayesian methods to study anisotropic models for
the distribution of gamma ray burst intensities and directions reported
in the {\it Third BATSE Catalog} (3B catalog) of gamma ray bursts.
We analyze data obtained using both the 64~ms and 1024~ms measuring timescales.
We study both purely local models in which burst sources (``bursters'')
are presumed to be distributed in extended halos about the Galaxy and
M31, and mixed models consisting of a cosmological population of
standard candle bursters and a local population distributed throughout
a standard Bahcall-Soneira dark matter halo with a 2~kpc core.
A companion paper studies isotropic models, including a variety of
cosmological models, using the same methodology adopted here, allowing
us to rigorously and quantitatively compare isotropic and anisotropic
models.  We find that the purely local models we have studied can
account for the 3B data as successfully as cosmological models, provided
one considers halos with core sizes significantly larger than those
used to model the distribution of dark matter.  A preference for
cosmological over local models, or vice versa, must therefore be
justified using information other than the distribution of burst directions
and intensities.  We infer core sizes for the halo distribution that
are smaller than one might expectbased on popular semiquantitative 
arguments that consider the superposed dipole moments of shells centered
on the Galactic center, and show why such arguments can lead to unwarranted 
conclusions.  We also find that the 3B data  do not constrain the width of 
power-law luminosity functions for burst sources.  This disagrees with the 
findings of previous studies; we elucidate the qualitative reasons for the lack 
of a constraint, and discuss why our results differ from those of earlier 
studies.  Our analysis of mixed models finds two families of models that can 
successfully account for the data:  models with up to 20\% of observed bursts
in a bright local population visible to $\sim 50$~kpc; and models with up to 
50\% of observed bursts in a dim local population visible only nearby (to
less than a disk scale height).  These models fit as well or better
than purely cosmological models.  They indicate that a surprisingly
large local, anisotropic component could be present whose size is
comparable to the sizes of hypothetical classes of bursts inferred
from analyses of temporal and spectral characteristics.  Finally,
as in our study of isotropic models, we find substantial systematic
differences between results based on 64~ms and 1024~ms data,
indicating that a thorough understanding of the distribution of
burst intensities and directions is likely to require detailed
analysis of temporal properties.
\end{abstract}

\keywords{Gamma rays: bursts}


\section{Introduction}

At the time of the launch of the {\it Compton Gamma Ray Observatory} ({\it 
CGRO}), the prevailing viewpoint among gamma ray astrophysicists was
that gamma ray bursts originate in the vicinity of neutron stars distributed
throughout the Galactic disk (see, e.g., the reviews
of Liang and Petrosian 1985, and Higdon and Lingenfelter 1990).
Perhaps the strongest evidence supporting this hypothesis was the
detection of absorption-like features at energies of $\sim10$--50~keV in
the spectra of bursts.  The KONUS experiment provided the earliest
evidence for the presence of such features (Mazets et al.\ 1981, 1982),
but the most conclusive evidence came from instruments on the 
{\it Ginga} spacecraft
(Murakami et al.\ 1988; Fenimore et al.\ 1988).  The features observed
by {\it Ginga} have high statistical significance, and can be
well-modelled as being due to cyclotron scattering in a strongly
magnetized plasma with field strength $B\sim 10^{12}$~G
(Wang et al.\ 1989; Lamb et al.\ 1989).  This field strength is typical
of that associated with both rotation-driven and accretion-driven
pulsars, suggesting that bursts are associated with strongly magnetized
neutron stars.  Further, requiring that a gravitationally
confined scattering region be static
implied that the burst sources so far observed were no further than a few 
hundred parsecs distant, otherwise the sources would have to be so luminous
that radiation presure would drive the scattering plasma away from the
source (Lamb, Wang, and Wasserman 1990; magnetic confinement may relax
this constraint).

The hypothesis that bursters formed a disk population 
seemed consistent with the most direct information
available about the spatial distribution of bursters: the distribution
of burst directions and intensities.
The apparent isotropy of the distribution of directions to bursts
(Atteia et al.\ 1987; Golenetskii 1988; Hartmann and Epstein 1989;
Hartmann and Blumenthal 1989) implied that we could be observing members of
a disk population only to distances smaller than a disk scale height---the
distance scale implied by magnetized neutron star models with a static
cyclotron
scattering region.  Reconciling a disk population with the cumulative distribution of burst intensities (the ``size-frequency'' 
distribution) was somewhat more problematic. 
The cumulative distribution of burst fluences, 
$S$, or peak energy fluxes, $F$, was
significantly flatter than the $-3/2$ power law expected
from sampling a spatial distribution from well within its characteristic
length scale,
but it appeared that selection biases could account for much of
the flattening (Yamagami and Nishimura 1986; Higdon and Lingenfelter 1986;
Mazets and Golenetskii 1987; Paczynski and Long 1988;
Schmidt, Higdon, and Hueter 1988).
The burst peak count rate, $C$, was proposed
 as a less
``biased'' intensity measure than fluence or peak energy flux, and
the distribution of peak count rates appeared to be consistent with
a $-3/2$ power law.  
Dispute remained over whether the
distribution began to flatten for the dimmest bursts (Jennings 1988),
but the number of faint bursts was too small to ascertain whether such
bursts were anisotropically distributed, as one would expect if the
faint bursts were observed from beyond a disk scale height.
Hartmann, Epstein, and Woosley (1990) modelled the distribution of neutron
stars in the Galaxy, and found the direction and
intensity observations to be consistent with
an association of bursts with Population I neutron stars, provided the
distribution was sampled to distances beyond $\sim 150$~pc, but no
greater than $\sim 2$~kpc.

Although the Galactic neutron star scenario appeared consistent with
the observations,
some investigators argued in favor of a cosmological origin for bursts
(e.g., Usov and Chibisov 1975; van den Bergh 1983; Paczynski 1986;
Goodman 1986).  The combination of isotropy of burst directions
and inhomogeneity implied by burst intensities is a natural
characteristic of such models, provided the observations sample
sources well beyond the local supercluster (Hartmann and Blumenthal 1989).
An additional motivation for considering cosmological models was
provided by Paczynski (1990) who, using
a different model for the distribution of old neutron stars than that
adopted by Hartmann, Epstein, and Woosley (1990), found a neutron star origin
inconsistent with the distribution of burst directions and
intensities.  Together, the work of Paczynski and of Hartmann,
Epstein, and Woosley implied that
the consistency of the Galactic disk model with the observations depended
on uncertain details of the models, particularly in regard to the distribution
of birth velocities of pulsars and the detailed form of the Galactic
potential (Hartmann, Epstein, and Woosley 1990).

Six years prior to the launch of {\it CGRO}, Meegan, Fishman,
and Wilson (1985)
reported detection of a single burst by a sensitive balloon-borne
detector that should have seen $\sim 43$ bursts if the population of
burst sources was spatially uniform to the distance sampled by the
detector, conclusively demonstrating that the cumulative distribution
of intensities of
dim bursts was flatter than the homogeneous $-3/2$ power law.
This was thought to be consistent with the Galactic disk paradigm, 
provided that the detector was able to see bursts from sources more distant
than a disk scale height, hence detecting inhomogeneity in the
source distribution.  
It was thus predicted that the Burst and Transient
Source Experiment (BATSE) on board {\it CGRO} would find faint
bursts concentrated in the Galactic plane, finally providing
compelling evidence for the Galactic disk neutron star paradigm.

Within a year of the launch of CGRO, BATSE observations spectacularly refuted
the Galactic disk neutron star hypothesis (Meegan, et al.\ 1992).  The
observations confirmed the
inhomogeneity discovered with the earlier balloon observations: 
the cumulative distribution of the peak fluxes of BATSE
bursts roughly follows a $-1$ power law and definitively rejects the $-3/2$
power law expected for a homogeneous distribution.  
Yet the distribution of the directions to these bursts is consistent with 
isotropy, with no significant concentration of burst sources in the Galactic
plane.  These basic features---apparent isotropy, and inhomogeneity---have
been only more conclusively demonstrated by subsequent BATSE observations
(Fishman et al.\ 1992, 1996).  

Although the BATSE observations definitively rule out a Galactic disk
origin for bursts, there is considerable controversy over whether
the BATSE data favor cosmological models over local models that distribute
burst sources in a large Galactic halo or corona, rather than
in a disk population.  Large scale
isotropy and inhomogeneity are natural qualitative features of
cosmological models for burst sources, so the BATSE observations have
revitalized interest in such models.  One can construct halo or 
coronal models that are consistent with the data, but they have length
scales considerably larger than those for matter distributions known
to be associated with the Galaxy prior to the BATSE observations, further
fueling interest in cosmological models.  But in the last five years,
evidence has accumulated indicating that there may be a population
of high velocity neutron stars with unbound or marginally bound orbits,
very possibly forming a large Galactic corona
(Lyne and Lorimer 1994; Frail 1996).  Ironically, some of
this evidence has been provided by observations of Soft Gamma
Repeaters (SGRs; see, e.g., Rothschild 1996), another class of gamma ray transient
studied with BATSE.  This, combined with suggestive but so far
unconclusive evidence for burst repetition (which is difficult to
reconcile with most cosmological models) has revived interest in
Galactic models, and enlivened the controversy over whether the data
can discern between cosmological and local alternatives
(see, for example, Lamb 1995 and Paczy\'nski 1995).

This paper is the third in a series in which we apply the principles
of Bayesian inference to the problem of inferring the spatial and
energy distribution of burst sources from burst direction and intensity
data provided by BATSE.  In Paper~I (Loredo and Wasserman 1995) we 
described the methodology and compared it to other methods in use.  
In Paper~II
(Loredo and Wasserman 1996), a companion paper to this one, we apply
the method to isotropic models (including cosmological models), using
the data from the recently released {\it Third BATSE Catalog}
(Fishman et al.\ 1996; hereafter the 3B catalog).  In the present paper,
we use the method to study anisotropic models.  Our method is uniquely suited 
to the study of these models, because it is the only method presently
available that is capable of analyzing the distribution of burst
directions and intensities {\it jointly}.  Since all anisotropic
physical models so far proposed have an anisotropy whose characteristics
vary with burst intensity, only a method capable of analyzing the
joint distribution can fully evaluate these models.
In addition, since we are using the same method to study both
isotropic and anisotropic models, we are able to quantitatively compare
them.  The Bayesian tool for doing this---the odds ratio---includes
a factor that accounts for the size of the parameter spaces of models,
resulting in an ``Ockham's Razor'' that automatically and objectively
accounts for model complexity and parameter uncertainty in such comparisons.

We presume the reader to be familiar with the notation and methodology
described in Papers~I and II; \S\S~2 and 3 of Paper II summarize
this information.  The next section presents an analysis of halo
models whose burst rate density is
spherically symmetric about the Galactic center and falls off with
radius $r$ like $1/[1 + (r/r_c)^2]$, where $r_c$ is a core radius
parameter; we analyze models with and without a similar halo centered
on M31.  We study models with ``standard candle'' and power law
luminosity functions.
In \S~3, we analyze models that superpose two populations:
a standard candle halo population like those analyzed in \S~2, and a standard
candle cosmological population.  For these models, we set the core
size of the halo population equal to 2~kpc, the value inferred by
Bahcall and Soneira (1980) in their study of the Galactic rotation
curve.  This allows us to obtain precise,
model-dependent constraints on the fraction of bursts that could be
associated with a known local matter distribution.  
We find this fraction can be quite
high, even though the halo population has a relatively small core size.  
Finally, in \S~4 we
discuss some of the implications of the work reported here and in
Paper~II.

Before moving on to the results of our study, we note some
additional distinguishing
features of this study.  No study has yet been published that has
analyzed anisotropic models using the 3B data.  This most recent
BATSE catalog contains data for about twice as many bursts as the 
earlier 2B catalog (Fishman et al.\ 1994), 
so new analyses using this data are of obvious
importance.  In addition, we analyze data for two of the three
trigger timescales included in the catalog:  64~ms (the shortest)
and 1024~ms (the longest).  Previous studies used only one
timescale; most studies used the 1024~ms timescale.  As shown
in Paper~II, the shapes of distributions of 64~ms and 1024~ms peak
fluxes differ, and there is evidence that the additional structure
present in the 1024~ms data is due to peak dilution (incorrect
peak flux estimation when the peak duration is shorter than the
measurement timescale).  Thus it is important to analyze data from
different timescales to ascertain what features of one's inferences
are robust.  Our method can be generalized to include temporal
information about bursts, as outlined in Paper~I; but the required 
information
is absent from the 3B catalog, and such an analysis is beyond the
scope of the current investigation.

Finally, we are not aware of a single study that
calculated correct constraints for the unknown parameters of
Galactic models, even using earlier BATSE data.  
Rather than using standard tools for calculating
confidence regions, investigators instead computed goodness-of-fit
statistics on grids throughout parameter space, and used contours
of constant significance to constrain parameters
(see, e.g., Hakkila et al.\ 1994).  This is not a
correct parameter estimation technique, and the resulting ``significance
regions'' are of no use beyond determining whether the best-fit model
is acceptable.  
Since similar techniques have been used in other fields, we
devote Appendix A to a general discussion of the problems with this approach.
In particular, we apply it to a simple Gaussian estimation problem,
and show that the methodology adopted by earlier investigators leads
to grossly incorrect confidence regions, and that the error {\it grows}
with the size of the data set.  
By contrast, the Bayesian methodology produces probability densities for
parameters of Galactic models for the 3B data, from which rigorous
constraints on model parameters may be derived directly.


\section{Overview of Method}

As described in Paper I and summarized in \S~2 of Paper II, Bayesian
inference requires that a model fully specify the {\it differential
burst rate}, $dR/d\F d\drxn$---the burst rate per unit time, peak
flux, and steradian---as a function of peak flux $\F$ and direction 
$\drxn$.  For most physical models, including the halo models we
analyze here, we calculate the differential burst rate from the
{\it burst rate density}, $\dot n(\rvec,\lum;\params)$, which gives the
burst rate per unit time, volume, and peak luminosity for
bursts from position $\rvec$ with peak luminosity $\lum$.  
Most such models have unknown parameters, here denoted collectively
by $\params$.  Once
we have specified the burst rate density, we can straightforwardly
calculate the differential burst rate according to  
equation (2.1) in Paper~II.  Here and
throughout our papers, the term ``peak luminosity'' and the
symbol $\lum$ refer to the peak photon {\it number}
luminosity, not the more common energy luminosity, $L$.  Similarly,
``peak flux'' refers to the peak photon number flux, not to
the energy flux, $F$.  For convenience, we use these symbols to
refer to the luminosity or flux in the nominal detected energy range
of 50 to 300~keV.

We undertake two distinct and complementary statistical tasks:
estimating or constraining unknown parameters in a particular
burst rate density model; and comparing rival burst rate density
models.  Different calculations are required to address these
tasks.  Most previous analyses of the BATSE data did not distinguish
these tasks; we discuss some consequences of this in Appendix~A.
The principal quantity underlying the Bayesian approach to these
tasks is the likelihood function for the parameters $\params$ of a model,
$\like(\params)$.  This is just the probability for obtaining
the 3B data, presuming a particular model is true, with its parameters
equal to $\params$.  We discuss how to calculate $\like(\params)$ in
detail in Paper~I; we offer a summary of the necessary calculations 
in \S~2 of Paper~II.  

Once the likelihood is available, parameter
estimation proceeds by examining the posterior distribution
for the parameters: the normalized product of
$\like(\params)$ and a prior distribution for $\params$.  We quote
the mode of the posterior as the ``best-fit'' parameter point, and
we present contours of constant probably density bounding regions
containing a specified amount of integrated
probability (``credible regions'') as summaries of the constraints
the data impose on the parameter values.
We adopt the same conventions for prior
probabilities and plots of posterior densities as described in \S~2 of 
Paper~II; that is, we choose parameter axes (linear or logarithmic) so
that the prior density is constant, resulting in
posterior densities that are simply proportional to the likelihood
as a function of plotted parameters.

Model comparison proceeds by comparing the average likelihoods of
competing models (using the prior as the averaging weight over the
parameter space).  The ratio of the average likelihood of one
model to that of another is the {\it Bayes factor}, $B$; it is the odds
in favor of the former model over the latter (presuming they would
be given even odds in the absence of the data).  Where possible, we
also provide the results of an asymptotic frequentist 
maximum likelihood ratio test.  Such a test is possible only for
nested models; in contrast, the Bayes factor is appropriate for
comparing any models.  In addition, the averaging underlying the
Bayes factor accounts for the sizes of the parameter spaces of the
competing models, implementing a quantitative and objective ``Ockham's
Razor'' that tends to favor simpler models over more complicated
competitors.

Most burst rate density models can be written in the form
$$
\dot n(\rvec,\lum;\params) = A f(\rvec,\lum;\sparams),\eqn(nAf)
$$
where $A$ is an amplitude parameter and the remaining parameters,
$\sparams$, are shape parameters.  As we noted in Paper~I, this
separation is useful both because the shape and amplitude parameters
typically reflect different physics, and because the amplitude
parameter can be removed from the analysis {\it analytically}
when we want to focus attention on the shape parameters alone.
To make inferences about the shape parameters alone (taking into account
the uncertainty in the amplitude parameter), we calculate the
shape parameter likelihood function according to equation~(2.9) in
Paper~II.  To make inferences about all parameters, 
we calculate the full likelihood function according
to equation~(2.7) of Paper~II.

Before proceeding to the results, 
we remind the reader that in \S~3 of Paper II we note several
approximations we must make in order to calculate the likelihood
for the 3B data (these approximations
must be made in {\it any} rigorous analysis, not just a Bayesian analysis).
Some of those approximations have no effect on analyses of the isotropic
models that are the focus of Paper~II; these deserve spectial mention
in this work.  In particular, we note in Paper~II
that the systematic errors in the directions reported in
the 3B catalog are not well understood (Graziani and Lamb 1996).  
The calculations reported
here all use the reported systematic error (a constant error
of $1.6^\circ$, to be added in quadrature with the statistical error).
To test the robustness of our findings with respect to the size of
the systematic errors,
we have repeated several calculations with the systematic error increased
to the value reported in the 2B catalog ($4^\circ$).  The resulting
changes in the locations of credible regions are completely negligible.
This is presumably because uncertainties on these small angular scales
are not of great importance for constraining large angular scale 
anisotropy.  The precise size and geometry of the direction uncertainties
is more important for assessing models with structure on small angular
scales, such as models that allow burst sources to repeat (see, e.g.,
Luo, Loredo, and Wasserman 1996).  More troubling is the fact the
new direction algorithm used to produce the 3B catalog significantly 
changed the inferred directions of many of the bursts included
in the 2B catalog; but the peak flux estimates of these bursts have
not been recalculated using the new directions.  We have no way of
estimating the resulting systematic uncertainty in the peak flux
estimates, although we note in Paper~II some reasons to expect them
to be small.


\section{Halo Models}

\subsection{Standard Candle Models}

We begin by considering ``standard candle'' halo models, for which
we can write the burst rate density as
\beq
\dot n(\rvec,\lum) = \dot n(\rvec)\,\delta(\lum-\lum_h),
\eeq
where $\lum_h$ is the standard candle peak photon number luminosity.
We take the spatial dependence of the burst rate density 
to be spherically symmetric around the center of the host galaxy
(the Milky Way or M31), falling with galactocentric radius, $r$,
according to
\beq
\dot n(\rvec) =
     { \dot n_0 \over 1 + \left(r\over r_c\right)^2},\eqn(n-vs-r)
\eeq
where $r_c$ is the core radius, and $\dot n_0$ is the burst rate per
unit volume at $r=0$.  The rate density is thus roughly constant for
$r\lesssim r_c$, and falls like $r^{-2}$ beyond $r_c$.  This is the
form often used to model the distribution of dark matter; in these
applications, estimated core sizes are typically a few kiloparsecs
(Bahcall and Soneira 1980).
Some investigators refer to halos with core sizes significantly larger
than those of dark matter halos as ``coronas.''  This spherically
symmetric model, though widely used, is clearly unrealistic for
distances of order or larger than $\sim 300$~kpc (about half the
distance to M31).
This is because the dynamical timescales at these distances
are of order the Hubble time.  We expect the distribution of
matter at these distances to reflect initial conditions, and thus
do not consider it plausible that it exhibit the isotropy or
radial dependence of our halo models.
We note below when our inferences probe
this unphysical regime.

Our halo models have three parameters:  $r_c$ and $\lum_h$ determining
the shape of the differential burst rate calculated from 
$\dot n(\rvec,\lum)$, and $\dot n_0$ determining its amplitude.
As in Paper II, it proves convenient to replace some of these
parameters with dimensionless parameters (some observable properties
actually depend only on the values of the dimensionless parameters).
The observable properties of a Galactic halo depend on the distance
from Earth to the Galactic center, $r_0$.  
This distance is not known precisely but is generally found to
be 8 to 9~kpc.  We thus
write $r_c = \rho_c r_0$, replacing the core size parameter with
$\rho_c$.  We also replace $\lum_h$ with a dimensionless
parameter, $\nu_h$, defined so that
\beq
\lum_h = \nu_h 4\pi r_0^2 \Ffid,\eqn(nuh-def)
\eeq
where $\Ffid$ is a fiducial peak photon number flux value which we set equal
to 1~cm$^{-2}$~s$^{-1}$ (near the BATSE detection limit).
With this definition, $\nu_h^{1/2}$ is the distance at which a burst
source produces a burst with flux $\Ffid$, in units of $r_0$.  To simplify
interpretation of our results, when
plotting functions of $\nu_h$, we provide an axis labeled
with values of $r_{\rm fid}/r_0 = \nu_h^{1/2}$.

Several properties of the halo population depend only on the
dimensionless parameters.  But
to infer the central burst rate density, we must fix $r_0$.
We use the IAU recommended value of 8.5~kpc (recent estimates favor the
somewhat smaller value of $R_0=8.0\pm 0.5$~kpc; see Reid 1993).  
To adjust our inferred values of the central rate density to correspond
to other choices of $r_0$, one would simply replace $\dot n_0$ with
$\dot n_0 (r_0/8.5~{\rm kpc})^{-3}$.

To calculate the effects of a halo around M31, we place the center of
M31 at a Galactocentric radius of 
$r_A=78.82 r_0$, corresponding to $r_A=670$~kpc when 
$r_0=8.5$~kpc (Allen 1973; Reid 1993 notes more recent values are as high as
760~kpc).  We presume 
the M31 halo to have the same parameters
as the Galactic halo, and thus its inclusion does not affect
the number of parameters.  We use the M31 position reported in
the {\sl Third Reference Catalog of Bright Galaxies}
(de Vaucouleurs et al.\ 1991).

We number models consecutively from those studied in Paper~II, and thus
denote this dual halo model $M_9$.  

The best-fit parameter values from separate analyses of the 64~ms and 
1024~ms data appear in Table 1 with statistics comparing
the models with standard candle cosmological models for the same
data (model $M_4$ in Paper~II).  These statistics include the maximum likelihood
ratio, $R_{9,4}$, and the Bayes factor, $B_{9,4}$.  Since halo models and
cosmological models are not nested, no simple frequentist method
is available for comparing them (such as use of the asymptotic
$\chi^2_\nu$ distribution of $2\log R$).  
The Bayes factor is not
constrained to nested models, however, and permits an objective
comparison of the models.  
One could calculate
an approximate significance associated with the maximum likelihood
ratio using Monte Carlo simulations, but we did not undertake this
since we consider the Bayes factor to offer a more complete 
and accurate comparison of models.

We find that, for each data set,
the best-fit halo model makes the data slightly more probable
than the best-fit standard candle cosmological model ($R_{9,4}>1$).
But when account is taken of parameter uncertainty, the
cosmological models are somewhat favored ($B_{9,4}<1$).  
The 64~ms data favor cosmological models more strongly than
the 1024~ms data; but neither data set decisively prefers one
model to another.

To give some indication of how the anisotropy of these models
affects the quality of the best-fit model, we constructed artificial
isotropic models with $dR/d\F d\drxn = (1/4\pi) dR'/d\F$, where
\beq
{dR' \over d\F} = \int d\drxn \; {dR \over d\F d\drxn}, \eqn(fake-iso)
\eeq
The differential rate in the integral was that for the best-fit
halo model.  These models had likelihoods insignificantly larger
than those of the anisotropic halos from which they were constructed:
2.7 times larger for the 64~ms data, and 1.5 times larger for the
1024~ms data (for reference, recall that a frequentist likelihood
ratio test assigns 95\% significance to a preference for a model with an 
additional parameter if the likelihood is increased by a factor
of 7.4).  Thus anisotropy is playing little role in assessing these
models in the vicinity of the best-fit parameters.  It plays an
important role, however, in bounding the credible regions.

Figure~1 displays contours of the joint posterior density for the
shape parameters resulting from analyses of the 64~ms
(Fig.~1a) and 1024~ms (Fig.~1b) data.  Crosses indicate the
best-fit parameter points.
The best-fit core radius and luminosity for the 64~ms data are nearly
three and ten times larger than those for the 1024~ms data, respectively.
But the 95\% credible regions overlap significantly, so the
difference between the best-fit
values of the shape parameters exaggerates the discrepancy between the 
shapes of the differential rates required to model the data sets in the
context of these models. 

The vertical and horizontal dotted lines in Figure~1 indicate where
the core size or fiducial distance equals $r_A$.  Our models should
be considered physically implausible at scales of order half this
or larger, as noted above.  In particular, the second mode evident
in Figure~1a at large $\nu_h$ and $\rho_c$ corresponds to models
that effectively consist of a single gigantic halo enclosing both
the Galaxy and M31, a physically implausible model.  We could have
eliminated this behavior by making the prior vanish in these implausible
regions, but instead left the prior nonzero in order to display
the behavior of the likelihood function in various extreme parameter
regimes (discussed further below).

The credible regions based on the 1024~ms data are significantly narrower 
than those based on the 64~ms data.  This is due in 
part to the larger number of bursts observed on the 1024~ms timescale.
In addition, in Paper~II we showed that there is significant evidence
for steepening of the distribution of 1024~ms peak fluxes for
$\F \gtrsim 30$~cm$^{-2}$~s$^{-1}$.  Hence the fit to the 1024~ms
data will be improved if the bright bursts are observed from
distances $\lesssim r_c$ (where the burst distribution begins to resemble
a homogeneous distribution), and this may be partly responsible for
making the posterior calculated from the 1024~ms data narrower than
that calculated from the 64~ms data.
Only models with a more flexible flux
distribution (such as the broken power law models studied in Paper II)
reveal discrepancies between the two data sets.

The credible regions for both data sets
have several common features:  values of $r_c$ and $r_{\rm fid}$ smaller 
than $r_0$ are excluded, as are values $\gtrsim r_A/2$ (except for
the a priori implausible upper right region of Fig.~1a, mentioned above); 
and in
the allowed region of parameter space, the inferred core radius
and burst luminosity are strongly and positively correlated.
We can understand these features as follows
(Hakkila et al.\ 1994 also discuss some of this qualitative behavior).

First, consider models with small cores, $r_c \lesssim r_0$ (i.e.,
$\rho_c \lesssim 1$).  When
$r_{\rm fid}\gtrsim r_0$ (i.e., $\nu_h \gtrsim 1$), the 
observable bursts are significantly
concentrated toward the Galactic center; such models are excluded because
of the apparent isotropy of the distribution of 3B burst directions.
When $r_{\rm fid}\ll r_0$ (i.e., $\nu_h \ll 1$), 
the visible part of the halo instead appears
nearly isotropic and homogeneous.  Although the observed distribution
of directions is nearly isotropic,
the observed distribution of burst peak fluxes falls less quickly with
$\F$ than 
the $dR/d\F \propto \F^{-5/2}$ behavior predecited by a homogeneous
population (see, e.g., \S~4 of Paper~II), so these models, too, are excluded.
Finally, when $r_{\rm fid} \gg r_0$, the peak flux distribution is
dominated by the $1/r^2$ part of the halo, which results in 
$dR/d\F \propto \F^{-3/2}$.  The observed distribution falls more
quickly than this (see, e.g., \S~4 of Paper~II).
Thus all models with $r_c \lesssim r_0$ are excluded, accounting
for the empty lower region of Figure~1.

Now consider models with large cores, so that $r_c\gg r_0$ 
(i.e., $\rho_c \gg 1$).  
When $r_{\rm fid}\ll r_c$ (i.e., $\nu_h \ll \rho_c$), the
BATSE data sample the halo distribution from within its characteristic
length scale, producing a visible distribution that is nearly isotropic
and homogeneous.  As just discussed, the homogeneity of such a population
is inconsistent with the data, so these models (to the left of the 
credible regions in Figure~1) are excluded.  When instead
$r_{\rm fid} \gg r_c$, the observable bursts are approximately
sampled from a $1/r^2$ distribution.  The flux distribution
associated with such a radial distribution is $dR/d\F \propto \F^{-3/2}$.
This flux distribution
falls more rapidly than is indicated by the data, and
thus models to the right of the contours are excluded 
(see \S~4 of Paper~II).

Thus only models with $r_c \gg r_0$, and $r_{\rm fid} \gtrsim r_c$ are
viable, hence the strong positive correlation.  In fact, the
posterior is aligned parallel to the line $r_c = r_{\rm fid}$.

Finally, the constraint at jointly large values of $r_{\rm fid}$ and
$r_c$ is due to the M31 halo becoming too prominent in this part
of parameter space. 
The dotted lines in Figures 1a and 1b indicate where $r_c = r_A$ and
$r_{\rm fid} = r_A$, and intersect just beyond the credible regions
(for the 64~ms data, very large halos that are significantly larger
than $r_A$ are also formally viable, although they are physically
implausible).
To further demonstrate the importance of including M31 to constrain
this region of parameter space, we recalculated the joint posteriors
without the M31 halo.  Credible regions for these purely Galactic
models appear in Figure~2.  They do not close at large values of
$\nu_h$ and $r_c$.

The constraint on models with large luminosities
($r_{\rm fid} \gg r_c$) is the only constraint that arose due to the
distant $1/r^2$ falloff of the halo density.  If the falloff is
less rapid, the constraint is weakened.  To illustrate this,
we studied a model with $\dot n(r) \propto 1/(r+r_c)$.  For
such a model, when $r_c\gg r_0$ and $r_{\rm fid} \gg r_c$,
the flux distribution is approximately $dR/d\F \propto \F^{-2}$.
As we showed in our study of power law flux distributions in
\S~4 of Paper~II, a $\F^{-2}$ power law adequately describes
the distribution of 64~ms peak fluxes and of all but the brightest
1024~ms peak fluxes.  Thus we expect large luminosity models {\it not} to 
be ruled out for $1/(r+r_c)$ halo models.  In Figure~3, we
show contours of the joint posterior for $\nu_h$ and $\rho_c$
for such a model, based on the 1024~ms data.  As expected, the
credible regions have opened up at large $\nu_h$ compared to those
in Figure~1.  But we consider these models to be of only academic
interest; we know of no compelling physical argument that leads to
$\dot n(r) \propto 1/(r+r_c)$.

The 95\% credible regions for the shape parameters derived from
both data sets overlap considerably.  This may seem to imply that
the two data sets are consistent with each other, but such a
conclusion is unjustified without examining the entire parameter
space, including the amplitude parameter, $\dot n_0$, that we
integrated over to obtain the marginal distribution for the shape
parameters.  Figure~4a shows contours of the joint distributions for $\nu_h$
and $\dot n_0$, conditional on the best-fit values of $\rho_c$
for each data set.  The regions allowed by the two data sets lie
far from each other.  The two data sets imply quite different
best-fit values of $\rho_c$, however; Figure~4a thus displays
``slices'' of the three-dimensional posterior taken in different
planes for each data set.  Figure~4b instead shows contours of the
posteriors conditional on $\rho_c=10$, a value lying inside the 95\%
credible regions for each data set.  The credible regions now lie
closer together; in particular, were we to project or integrate the
distributions along either axis, they would overlap substantially.
However, Figure~4b makes it clear that the full posteriors are disjoint,
the nature of the discrepancy being that for any given values of
the luminosity and core size, the 64~ms data imply a burst rate
density over 50\% larger than that implied by the 1024~ms data.
We found similar behavior in our study of cosmological models
in Paper~II, and noted there that such behavior could easily arise
if bursts have peaks substantially briefer than 1024~ms, so that
the peaks are ``diluted'' when measured on this timescale.  Further
discussion appears in \S~6 of Paper~II.

\subsection{Models With Power Law Luminosity Functions}

In their study of Galactic halo models, Hakkila et al.\ (1995) concluded
that models with power law luminosity functions that span a dynamic
range greater than about 5 are incompatible with the BATSE data.  If
true, we would consider this an argument {\it against} halo models.
Given the extreme diversity of burst temporal behavior, and the broad
diversity of burst spectra, it seems highly unlikely to us that the
burst phenomenon has a single standard luminosity or a very narrow
luminosity function.  Thus we would consider incompatibility of broad
luminosity functions with the BATSE data to be evidence that the
success of standard candle halo models is fortuitous.  We have
therefore sought to verify the conclusion of Hakkila et al..
Our study is guided by our analysis of cosmological models with
broad luminosity function in Paper II.  There we found that the
constraint on cosmological luminosity functions reported by
Hakkila et al.\ (1994) was incorrect; they reported that power law
luminosity functions had to have a dynamic range less than about 10,
whereas we find the dynamic range to be completely unconstrained.
We gave a simple and intuitive explanation of the lack of a constraint
in \S~5.3 of Paper II.  Similar arguments should apply here, except
that the presence of M31 could possibly complicate matters.  In fact,
we find that for halo models, too, the dynamic range of a power law
luminosity function is not constrained by the data.

We now consider models with a bounded power law luminosity function,
so that in the interval $[\lum_l, \lum_u]$ the burst rate density is
\beq
\dot n(\rvec,\lum) = \dot n(\rvec)\,A \lum^{-p},\eqn(ndot-plh)
\eeq
with $\dot n(\rvec)$ given by equation \ceq{n-vs-r}, and $A$
a normalization constant determined by $p$, $\lum_l$, and $\lum_u$.
Outside of the luminosity interval, the burst rate density vanishes.

To calculate the differential burst rate implied by equation~\ceq{ndot-plh},
we must perform an integration over $\lum$, as well as the radial
integral required to calculate the differential rate for standard
candle models.  Since calculation of the likelihood function
requires numerous three-dimensional integrals of the differential
rate, the extra integration dimension makes analysis of models with
luminosity functions
computational burdensome.  We thus have not explored the full
parameter space of this model, and we cannot report a Bayes factor
for it.  The study of Hakkila et al.\ (1995) is similarly constrained.

We use the qualitative features of models with luminosity functions,
together with some of the findings of Paper II, to guide our restricted
search of parameter space.  In particular, we fix the power law index,
$p$, based on knowledge of the behavior of the flux distributions implied
by power law luminosity functions, and the the slope of the flux
distribution implied by the data and inferred in \S~4 of Paper II.
As we noted there, it is simple to show that a bounded population
that vanishes beyond radius $r_c$ implies a differential flux
distribution proportional to $\F^{-5/2}$ for bursts with fluxes brighter
than $\lum_u/(4\pi r_c^2)$, and proportional to $\F^{-p}$ for
dimmer bursts.  The flux distribution thus mimics the luminosity
function below the flux corresponding to observing the brightest
bursts at the boundary of the distribution.  The flux distribution
vanishes below $\lum_l/(4\pi r_c^2)$, the flux of the dimmest bursts
at the boundary.

Halo models complicate this picture in three ways:  the boundary is
not ``hard,'' the distribution falls like $1/r^2$ beyond $r_c$;
we are offset from the geometric center of the distribution, leading
to anisotropy; and the halo around M31 introduces further anisotropy
and changes the flux distribution for distant bursts.  
To guide our choice of parameters, we focus on 
the flux distribution, and consider the properties of a single halo
observed from its center.  A simple analytic calculation, described
in Appendix B, reveals
the generic behavior plotted in Figure~5 for the parameter range
of interest in studying bursts.  At large fluxes, $dR/d\F \propto \F^{-5/2}$,
the familiar power law associated with a homogenous population.
Going to lower fluxes, the differential rate first flattens to
$dR/d\F\propto \F^{-p}$ at a flux of $\lum_u/4\pi r_c^2$, 
and then flattens further to 
$dR/d\F\propto \F^{-3/2}$ below a flux of $\lum_l/4\pi r_c^2$.  
The dynamic range for the $\F^{-p}$ behavior
is $\lum_u/\lum_l$.

As discussed in Paper II, there is no evidence that BATSE has yet
observed the $\F^{-5/2}$ behavior expected for bright bursts (although
there may be evidence for much more significant steepening in the 1024~ms 
data).  In the context of models with luminosity functions, the
observed data must span part of the $\F^{-p}$ and $\F^{-3/2}$ regions.
In this case, the simple calculation that produced Figure~5
indicates that the dynamic range of a luminosity function with 
index $p$ can be constrained only if the differential flux 
distribution flattens from $\F^{-p}$ to $\F^{-3/2}$ over the range of 
the data.  No such flattening is visible to the eye in the ``complete''
portion of the data (above $\F = 3$~cm$^{-2}$~s$^{-1}$ for the 64~ms data,
or $\F = 1.5$~cm$^{-2}$~s$^{-1}$ for the 64~ms data).  But it remains
possible that the flattening due to the efficiency function is
hiding some intrinsic flattening, or that the complications due to
our offset from the Galactic center and the presence of M31
produce characteristics in the joint distribution of fluxes and
directions that yeild a constraint.

Accordingly, we investigated models with $p=1.8$, and with a dynamic
range of $10^3$.  We chose $p=1.8$ because we found that both phenomenological
and cosmological models with $\F^{-1.8}$ low-flux behavior fit both
data sets well in Paper II.  The dynamic range of $10^3$ is many
times larger than the upper limit of 5 found by Hakkila et al., and
is a value that we showed to be consistent with the data in the
context of the isotropic models studied in Paper II.  With these
parameters fixed, only two shape parameters remain:  the core
radius parameter, $\rho_c$, and the upper limit of the luminosity
function, $\lum_u$.  As with standard candle models, we introduce
a dimensionless parameter, $\nu_u$, for the luminosity, defined
so that
\beq
\lum_u = \nu_u 4\pi r_c^2 \Ffid.\eqn(nuu-def)
\eeq

Figure~6a shows plots of the logarithm of the shape parameter
likelihood as a function
of $\nu_u$, when $\rho_c$ is fixed at the most probable value
found from the previous analysis of {\it standard candle} models
($\rho_c = 18.1$ for the 64~ms data; $\rho_c = 7.8$ for the 1024~ms data).
Curves are shown both for the 64~ms (solid) and 1024~ms (dashed) data.
The likelihoods have been divided by the maximum likelihoods for
the corresponding best-fit standard candle models.  For the
64~ms data, models with $\nu_u \approx 7000$ are actually slightly
superior to standard candle models, so it is clear that the 64~ms data
allows luminosity functions with a dynamic range of at least $10^3$.
On the other hand, the best model for the 1024~ms data has a
likelihood almost 40 times lower than that of the best standard
candle model for these data.  This would seem to indicate a
reasonably strong preference for narrow luminosity functions for
these data.  In fact, Hakkila et al.\ (1995) studied
1024~ms data (although they used {\it energy} flux data, not
photon number flux data).

However, Figure~6a constrains $\rho_c$ to its standard candle best-fit
value.  Although we have not performed a full study of the
$(\rho_c,\nu_h)$ parameter space, Figure~6b presents results similar
to those shown in Figure~6a, but with slightly reduced values of
$\rho_c$ ($\rho_c=10$ for the 64~ms data; $\rho_c=5$ for the 1024~ms data).
The likelihoods for both 64~ms and 1024~ms models have improved.
In fact, the likelihood for the best 1024~ms model is now only 2.7
times smaller than that of the best standard candle model, and is
clearly acceptable.  It is likely that futher adjustment of
$\rho_c$ or of $p$ would improve the likelihood further.  Thus
models with {\it very} broad luminosity functions are entirely
consistent with the data, and are competitive with both standard candle
halo models and cosmological models.

Hakkila et al.\ (1995) offer very few details about their study, but
do note that they explored luminosity function models only with
parameters in the vicinity of the best-fit standard candle models.
The core radii for the acceptable models in Figure~6b are lower than
those of acceptable standard candle models, and the luminosity upper
limits are significantly higher than the luminosities of acceptable
standard candle models.  We suspect that an unnecessarily restrictive
study of the parameter space is responsible for the different conclusions of
Hakkila et al.\ (1995).

\subsection{The Anisotropy of Halo Models}

Some of our findings may appear somewhat surprising in the light of
simple arguments regarding the anisotropy of halo models.  
Hartmann (1994) and Hartmann et al.\ (1994) point out that a Galactocentric 
shell of radius $r$ has a dipole moment $D=\langle \cos\theta\rangle$
given by
\beq
D(r) = {2 \over 3}\, {1 \over \rho},\eqn(D-Hart)
\eeq
where $\rho = r/r_0$.
Arguing that any spherically symmetric halo model may be thought of
as the weighted sum of a series of such shells, they use
equation~\ceq{D-Hart}\ to constrain the length scale of halo models.
Briggs et al.\ (1994) use this result, finding that shells with $r<110$~kpc
have dipole moments inconsistent with the dipole moment of the
proprietary sample of 1005 bursts that they analyze.  They argue that
any bursts interior to this radius must be balanced by those exterior
to it, implying that only models with very large length scales will be 
acceptable.  Whether
the argument is meant to constrain the core size or the distance to
the faintest sources is not clear.

In contrast to this argument, we find that 
models with core sizes as small as $2r_0= 
17$~kpc and with bursts visible only out to $\approx 50$~kpc lie
within the 95\% credible region for standard candle models for the
1024~ms data; both scales are substantially smaller than the
$\sim 100$~kpc scale identified by the Galactocentric shell 
argument.  We are analyzing a smaller data set than is available to
Briggs et al. (1996); but their own analysis of specific models also
identified acceptable models with length scales very much smaller
than the $\approx 100$~kpc scale.  The Galactocentric shell 
argument has been so 
widely cited that these results certainly deserve some comment.

In fact, this argument is misleading because it fails
to distinguish the artificial anisotropy arising from
the displacement of the centers of Galactocentric shells from the observing 
point (the Sun), and the actual, intrinsic anisotropy best quantified by 
calculating angular moments of shells centered on the Sun.
The simplest way to see this is to consider observations of a
completely homogeneous population.  Of course, the observed dipole
moment of such a population has an expectation value of zero, not only
for the entire data set, but also for every constant-$\F$ subset.
In this case, the dipole moments of the Galactocentric shells considered 
in the argument of Hartmann et al.\ (1994) are entirely an artifact of their
offset from the observing point.
A vanishing moment cannot be realized by any positively weighted
superposition of $D$ values given by equation~\ceq{D-Hart},
but certainly a proper treatment of the observation of such shells
must produce a vanishing dipole moment.
The way this comes about is elucidated by Figure~7, which shows
cross sections of a large-$r$ Galactocentric shell, and of a 
Sun-centered shell which we take to have nearly the radius at which the
dimmest observable bursts lie.  Near the observing limit, the
dipole moments of the {\it observable} part of Galactocentric
shells are {\it negative}, because only those parts of the shells in the
direction of the anticenter (highlighted) are visible.  These
negative contributions cancel the positive moments of smaller shells
that are completely visible, so the total dipole moment vanishes.
This effect is ignored
in the argument of Hartmann et al.\ (1994) and Hartmann (1994).

Hartmann et al.\ (1994) sought to constrain the length scale of the
burst population in a manner that was both model-independent and
avoided complications due to observational selection effects.
Their focus on Galactocentric shells achieved model independence in that
all models that are spherically symmetric
about the Galactic center can be built by superposing such shells.  
But the offset of the shells from the Sun
makes consideration of selection effects crucial for calculating
observed moments, and seriously complicates any discussion of the
intensity dependence of the expected anisotropy.  A model will predict
anisotropy in the burst data as a whole only if shells centered on the Sun 
exhibit anisotropy.  Unfortunately, the anisotropy of such shells depends
on how the burst rate density varies with Galactocentric radius.
We therefore find it difficult to see how one could actually constrain
length scales of spherically symmetric halo models in a model-independent
manner.

If we forsake model independence, we can gain some
insight into why halo models with distance scales smaller than $\sim 100$~kpc
are acceptable by calculating the dipole moments of shells of radius
$r'$ from the Sun, assuming the standard halo profile used above.
A straightforward analytic calculation, described in Appendix C,
shows that a good approximation is
\beq
D(r') \approx {2 \over 3}\, {1 \over \rho' + \rho_c + {1\over \rho'}},\eqn(D-rp)
\eeq
where $\rho' = r'/r_0$ and $\rho_c = r_c/r_0$.  Comparing with
equation~\ceq{D-Hart}, we see that a Solar-centric shell of a given
size has a dipole moment that is always smaller than that of
a Galactocentric shell of the same size.  Also, in stark contrast to
the Galactocentric case, Sun-centered shells with
small $r'$ have {\it small} dipole moments, because they
are observed across a region over which the distribution appears
homogeneous.  The dipole moment takes its maximum value for
$\rho' \approx \sqrt{\rho_c + 1}$; small observed moments can result
from populations observed predominantly at distances {\it smaller} than this,
as well as from those with many distant bursts.
Finally, in contrast to equation~\ceq{D-Hart}, 
equation~\ceq{D-rp}\ clearly displays the distinct roles played by the
two distance scales: the observing distance, $r'$, and the core size,
$r_c$.  These characteristics of $D(r')$ make feasible the
acceptability of models with distance scales smaller than 100~kpc;
but only a rigorous, model-dependent statistical analysis can produce 
quantitative constraints.

It may also appear surprising that the likelihood of models with luminosity
functions is
improved by {\it decreasing} the core radius below that of standard
candle models, presumably increasing
their anisotropy.  Such behavior was noted in passing by Brainerd
(1992), but left unexplained.  Figures~8 and 9 elucidate the reason
for this somewhat counterintuitive behavior.  Figure~8a shows
the burst rate (per unit distance) as a function of radius from
the Sun, $r'$, for the
standard candle model that best fits the 1024~ms data.  The burst
rate per unit distance was calculated according to
\beq
{dR \over dr'} = r'^2 \int d\drxn \, \dot n(\rvec) 
   \int d\lum\, f(\lum) \bar \eta\left({\lum\over 4\pi r'^2},\drxn\right),
    \eqn(dRdr-def)
\eeq
where $f(\lum)$ is the luminosity function (a $\delta$-function for
standard candle models), and $\bar\eta(\F,\drxn)$ is the time-averaged
efficiency for detecting bursts of flux $\F$ from direction $\drxn$.  
As a simple qualitative measure of the
anisotropy of the burst population, Figure~8b shows the dipole moment,
$D$ (the average of the cosine of the angle between a burst and the unit
vector to the
Galactic center), of a shell of a given radius.  It is slightly
negative at small distances due to the slightly negative dipole moment
of the BATSE sky exposure map; it is negative at large distances due
to M31, which is in the hemisphere opposite  of the Galactic center.
Figures~8c and 8d show similar results for the best broad
luminosity function model we studied for the 1024~ms data, a model with
$p=1.8$, $\rho_c=5$, $\nu_u=2\times10^3$, and a dynamic range
of $10^3$.   As expected, according to this model the bursts are
observed from smaller distances than in the best standard candle model,
and have significantly larger anisotropy on constant-radius shells.

In contrast, Figure~9 shows the dipole moments for these models
as functions of {\it peak flux} rather than of radius; Figure~9a
is for the standard candle model and Figure~9b is for
the model with a broad luminosity function.  Also shown are the
dipole moments (with statistical error bars) of the best-fit burst
positions for bursts in five
flux bins of equal logarithmic width.  Although the
luminosity function model exhibits a larger dipole moment than the
standard candle model at all fluxes, its dipole moment is significantly
lower at all fluxes than its maximum value in the $D$ vs.\ $r/r_0$
plot of Figure~8d.  Further, it peaks at larger fluxes than does
the standard candle curve.  There are fewer bursts at these fluxes
than at lower fluxes, and thus weaker constraints on anisotropy, as
the error bars illustrate (the data also slightly favor
large dipole moments at large fluxes).
Due to the broad luminosity function, the bursts from a shell with
a large dipole moment appear at a wide variety of fluxes, mixed with
bursts seen from other shells with less anisotropy.  The luminosity
function thus ``spreads'' the anisotropy out over a large range
of fluxes, allowing populations with smaller core radii to fit
the data.


\section{Two-Population Models}

\subsection{Motivation}

Many authors have noted the bewildering variety of characteristics
exhibited by GRBs, particularly in regard to their temporal behavior.
Some investigators have regarded this variety as evidence that GRBs 
are not due to a single underlying phenomenon, and this possibility has
motivated many attempts to classify bursts according to their
temporal or spectral properties.  Several such attempts predate
the BATSE observations, but no consensus was reached on
the number or defintions of burst classes (see the reviews of Hurley 1986
and Higdon and Lingenfelter 1990 for further discussion).  

The large and uniform data base provided by BATSE offers new opportunities for
burst classification.  The most intriguing finding so far reported is
the discovery of suggestive evidence for two classes of bursts distinguished
by light curve morphology:  Lamb, Graziani, and Smith (1993) identify two
classes based on a simple measure of light curve variability; Kouveliotou 
et al.\ (1993) instead identify two classes based on simple measures of burst
duration.  The statistical significances of these discoveries are
difficult to determine, due to both ambiguity in modeling the
classes and the possibility of subtle effects mimicking the signatures
of distinct classes for certain measures of light curve morphology
(see, e.g., Wang 1996).  Both groups of investigators find that class
membership is correlated with other burst properties (such as hardness
or brightness), and some recent studies find supporting evidence in
larger data sets (Kouveliotou et al.\ 1996; Katz and Canel 1996).
This may support the reality of the classes, although a correlation between
temporal and other properties could arise from a single underlying
mechanism.

Two somewhat more controversial discoveries also suggest that there may
be two or more classes of bursts.
As noted in \S~1, the {\it Ginga} experiment discovered strong evidence
for the presence of low-energy absorption-like features in about 20\%
of the bursts it observed.  Such bursts may constitute a distinct class,
and in the best-studied models for the formation of the features---the
cyclotron scattering model mentioned in the Introduction---the sources 
of these bursts would have 
to lie within several hundred parsecs of the Sun if the scattering
region is gravitationally confined.  The distribution of
burst intensities is not compatible with such a distribution, so bursts
without features would presumably have a very different (perhaps
cosmological) spatial distribution.
BATSE has yet to confirm the presence of
such features in bursts, but the present limits and candidate
features are not inconsistent
with the {\it Ginga} results (Palmer et al.\ 1994; Band et al.\ 1994, 1995;
Briggs et al.\ 1996),
and there are some questions regarding how well BATSE could detect
such features (Band et al.\ 1995; Paciesas et al.\ 1996; Freeman et al.\ 1996).
The small number of bright bursts observed
by {\it Ginga} results in large uncertainty in the inferred fraction of
bursts with such features, but it is interestingly similar to the size
of the smaller of the two classes inferred from light curve morphology.
In addition, several investigators reported suggestive evidence for
burst repetition based on clustering of the burst directions reported
in the 2B catalog (Lamb; Wang and Lingenfelter).  Only a small subset
of bursts exhibited clustering; these may comprise a distinct class.
The evidence for repetition is
absent in the 3B catalog (Hartmann et al.\ 1996; Luo, Loredo, and
Wasserman 1996; Luo and Wasserman 1996), but the directions reported in
this catalog were calculated with a new burst direction algorithm that
has systematic errors that appear to be significantly more complicated
than those in the 2B catalog (Graziani and Lamb 1996); the additional
complications have yet to be fully characterized or incorporated in 
searches for evidence of burst repetition.

Although none of the observational studies just described is completely 
compelling,
together they suggest that one take seriously the possibility that
bursts arise from two or more populations with distinct characteristics.
Some theoretical studies also suggest that bursts might be produced
by distinct phenomena that are observable at different spatial scales.
For example, Wasserman and Salpeter (1995) have suggested that the universal
behavior exhibited by the rotation curves of galaxies may indicate the
presence of a baryonic halo around galaxies consisting of stellar
remnants of various types resulting from the evolution of a population of
stars produced prior to the collapse of the protogalaxy. 
They have calculated various collision rates for these
remnants, and they identify two types of collisions that are energetically
plausible sources of bursts and that might produce
bursts at roughly the rate observed.  Collisions between neutron stars
and asteroids could produce bursts observable throughout the halo of
the Galaxy at rates $\sim 10^2$~yr$^{-1}$.  Collisions between
pairs of neutron stars (not in binary systems) could produce bursts 
observable to cosmological
distances at a similar rate.  In this picture, bursts thus arise from
both a local halo population and a cosmological population of sources.
Katz (1996) envisions a scenario that similarly divides burst sources
into local and cosmological components, the former due to magnetic
reconnection in neutron star magnetospheres, and the latter due to
the interaction of a fireball (possibly produced by a collision between a
neutron star and a companion neutron star or black hole) with surrounding
clouds, as suggested by Shemi and Piran (1990) and M\'esz\'aros
and Rees (1993).

These observational and theoretical arguments have motivated us to
study two-population models for burst sources comprised of a local
halo population and a cosmological population.  
More pragmatically, such models
are interesting to study simply because they
provide a convenient framework in which we can place precise (but
model-dependent) constraints on the fraction of bursts in a local,
anisotropic population.  Of course, since
we have already demonstrated that pure halo models and cosmological
models can independently account for the data equally well, 
we know that one can create successful two-population models
with any desired fraction of bursts in the local population.  However,
here we restrict the halo population to reside in
{\it Bahcall-Soneira halos
with a 2~kpc core}.  These halos have $\rho_c = 0.235$, and are
thus {\it not} acceptable by themselves.  However, it is widely believed
that a dark matter halo with a core size similar to this exists, on the
basis of Galactic rotation curve measurements.  Further, this halo
scale is similar to that of the baryonic halos studied by Wasserman
and Salpeter (1995).  Thus, unlike the halos of the previous section,
the halos we study here are not created solely for the purpose
of hosting bursts.

Finally, these models are also of interest in that they offer
a model-dependent test of the adequacy of cosmological models:
even if one rejects these models as subjectively implausible a priori,
if models with a substantial halo component have higher likelihoods
than purely cosmological models, this could indicate that the data
prefer more complicated alternatives to the cosmological
models we studied in Paper~II.  Whether the improvement is due to
slight anisotropy or improved fitting of the flux distribution could
offer clues as to which alternatives one might study.

\subsection{Model Specification}

For the cosmological component, we consider a distribution
of standard-candle burster sources in a $\Lambda=0$, $\Omega_0=1$ universe. 
This model is discussed in detail in Paper II (Model $M_4$ of \S~5.1). 
This model component has a single shape parameter, 
the standard candle photon number luminosity, $\lum_c$.  We write 
$\lum_c$ in terms of a dimensionless luminosity, $\nu_c$, according to
\beq
\lum_c = \nu_c (4\pi c^2 / H_0^2)\F_{\rm fid}  K(0),\eqn(nuc-def)
\eeq
where $H_0$ is
Hubble's constant, $K(z)$ is a redshift-dependent spectral correction
function (similar to a ``$K$-correction''), and
$\F_{\rm fid}$ is a fiducial value of the observed flux, which
we set equal to 1 cm$^{-2}$s$^{-1}$.  For $\nu_c=1$, 
$\lum_c \approx 10^{57}$s$^{-1}$, corresponding to a luminosity of
approximately $10^{51}$ erg s$^{-1}$ for $H_0=100$ km s$^{-1}$,
and burst sources with $z\approx 1$ for bursts with $\F\approx 1$.
The best-fit value of $\nu_c$ in a purely cosmological model is
$\nu_c \approx 0.4$ for both the 64~ms and 1024~ms data.
We note that cosmological models with density evolution or with
luminosity functions studied in Paper~II do not substantially improve
on standard candel models.  Thus although this model is simple, it
is representative of all successful cosmological models for the 
3B data.

For the local component, we consider a dark matter
halo distribution of standard-candle sources with the core radius fixed at
$r_c=0.235 r_0 = 2$~kpc.  As noted above,
this choice thus corresponds to associating local bursters
with a known distribution of matter in the Galaxy.  A single shape
parameter describes the flux distribution resulting from such a
population:  the standard candle photon number luminosity, $\lum_h$.
As in the previous section, we write this luminosity in terms of a 
dimensionless parameter, $\nu_h$, according to equation~\ceq{nuh-def}.
As shown below, the only tenable models have values of $\nu_h$ small
enough that the contribution of an M31 halo would be completely
negligible.  We thus omitted the M31 halo from the calculations, to
speed them up.

The full two-component model combines the cosmological and halo
populations so that a fraction, $f$, of the observable bursts comes
from the halo population.  The model thus has three shape parameters,
the two luminosity parameters, $\nu_c$ and $\nu_h$, and the halo
fraction, $f$.  Note that since $f$ parametrizes the fraction of
{\it observable} bursts in the halo component, its meaning depends on
the instrument providing the data being analyzed.  In particular, its
meaning is different for the 64~ms and 1024~ms timescale data sets.

\subsection{Results}

Figure~10 shows the profile likelihood as a function of the halo
luminosity, $\nu_h$.  The profile likelihood for a subset of a model's
parameters is simply the likelihood maximized over the remaining
parameters; it is sometimes a good approximation to the marginal
distribution for the subset of parameters.  Here we use it simply
as a convenient display of some of the features of the full joint 
posterior.  The profile likelihood reveals the posterior to be
multimodal, with (at least) one mode
in the region where $\nu_h\gg 1$ (corresponding to sources more distant
than the Galactic center)
and another where $\nu_h\ll 1$ (sources much closer than the Galactic
center).  We thus examine models with dim ($\nu_h<1$) and 
luminous ($\nu_h>1$) halo burst sources separately.
The 1024~ms data imply a third small mode for $\nu_h\approx 1$; such
solutions are inconsistent with the 64~ms data, although we do discuss
them briefly below.

\subsubsection{Models With Luminous Local Bursters}

Table~2 presents the best-fit parameters for models with luminous
halo bursters based on both data sets.  Also listed are
the ratios of the best-fit likelihoods to those of the best-fit
standard candle cosmological models (corresponding to $f=0$), and
the Bayes factors favoring two-population models with $1<\nu_h<10^3$
over standard candle cosmological models.  The Bayes factors were
calculated with a relatively crude grid in $\nu_h$, and are probably
accurate to only one significant figure.

The Bayes factors are not significantly different from unity,
indicating no strong preference for or against these models over
purely cosmological models, once account is taken of the unknown
parameters.  Since purely cosmological
models are nested within our two-component models, we can also
easily calculate calculate the significance associated with a frequentist likelihood ratio test (this was not possible for the pure halo
models studied above).  Table~2 lists the probability $p(>R)$ of seeing a 
larger maximum
likelihood ratio, presuming the best-fit cosmological model
is true).  The table lists approximate significances, calculated
using the asymptotic $\chi_2^2$ distribution of $2\log R$.  
This approximate frequentist test agrees with the Bayes factor
in the sense that it does not find the preference for two-component
models to be very significant.  The Bayes factor is a more complete
comparison; it accounts for the sizes of the parameter spaces of
the models, comparing the {\it average} rather than maximum likelihoods,
and thus implementing an automatic and objective ``Ockham's Razor.'' 

The best-fit halo luminosities imply a local population visible
to $\approx 50$~kpc ($r_{\rm fid} = 49.8$ and 37.1~kpc for the 
64~ms and 1024~ms best-fit parameters).  The cosmological luminosities
are significantly smaller than unity, implying a cosmological
component that is nearly homogeneous over the span of the 3B~data.
These $\nu_c$ values are significantly smaller than the values
of order unity favored in purely cosmological models.

Figure~11 shows contours of the joint distribution for the observable halo
fraction, $f$, and the cosmological
luminosity, $\nu_c$, conditional on the best-fit values of $\nu_h$
for each data set (listed in Table~2).  (The joint marginal distributions
will be somewhat wider after averaging over $\nu_h$, but not significantly
so because correlations with $\nu_h$ are weak.)  Halo fractions of
order 10\% are favored, and surprisingly large values of $f$ can be 
tolerated:
models with $f\approx 0.2$ lie within the 95.4\% credible regions.
Thus halo fractions consistent with the sizes of the burst classes
hypothesized in the studies reviewed above are consistent with the
3B data.

We constructed isotropic versions of the best-fit models in the
manner discussed above (see eqn.~\ceq{fake-iso}) and calculated their
likelihoods.  The isotropic versions had likelihoods that were very
slightly larger
(factors of 1.22 and 1.12 larger for the 64~ms and 1024~ms data,
respectively).  The anisotropy of these models is playing little
role in determining the quality of the best-fit model.   We can
thus understand the best-fit models by focusing on their predicted
intensity distributions.

In Figure~12 we show the cumulative distributions 
of burst fluxes predicted by the best-fit models, along with the observed
cumulative histograms.  The dashed curve (associated with the right axis)
shows the negative logarithmic slope of the cumulative distribution.
These models are more successful than
purely cosmological models because the halo component allows the
differential rate to steepen more quickly with flux.  Figure~13 shows
the total differential rate (solid) and
the separate differential rates for the cosmological (dotted) and halo
(dashed)
components of the best-fit models, illustrating how the steepening
comes about.  At low fluxes, the halo component differential rate 
is $\propto\F^{-3/2}$ and is small compared to the cosmological component.
But above a flux of 10 to 20~cm$^{-2}$~s$^{-1}$, the halo component
quickly steepens to a $\F^{-5/2}$ power law, and is comparable in
magnitude to the cosmological component.  In this manner, the slowly
flattening cosmological component accounts for the $\F^{-2}$ behavior
exhibited by dim bursts, and both components together account for
the bright bursts, whose rate falls off more steeply.  The halo
component not only makes the change to $\F^{-3/2}$ behavior more
abrupt; it also somewhat enhances the number of bursts at large
fluxes.  This enhancement appears as a small bump in the total differential
rate in Figure~13; it is not apparent in the cumulative distribution.

\subsubsection{Models With Dim Local Bursters}

Table~2 also lists the best-fit parameters for models with dim ($\nu_h<1$)
halo bursters based on both data sets, along with measures of the
quality of fit of these models.  For the 64~ms data, the likelihood
is maximized for the smallest value of $\nu_h$ we examined, 
$\nu_h = 10^{-3}$.  For
the 1024~ms data, there is a definite mode at $\nu_h\approx 10^{-2}$, 
but the likelihood remains large for $\nu_h = 10^{-3}$.  In both cases,
the favored models have bursts visible only from within a kiloparsec
from the Sun.  The local component is thus nearly isotropic and 
homogeneous.  The best-fit cosmological luminosity is significantly
larger in these models than in models with a luminous local population.
The cosmological component thus exhibits strong apparent inhomogeneity
over the span of the 3B data.

The maximum likelihood is much larger than that of purely cosmological
models; but again, the Bayes factor indicates that these models are
neither decisively favored nor disfavored compared to simpler
purely cosmological alternatives.  We constructed isotropic versions
of the best-fit models by averaging over direction, as described above.  
The likelihood of the isotropic version of the 64~ms best-fit model
was greater than that of the actual anisotropic model by a small factor
(4.3).  The likelihood of the 1024~ms isotropic model is 2.5 times
{\it smaller} than that of the anisotropic model.  The small anisotropy
of the local component actually helps this model, though by an
insignificantly small amount.  This is an interesting result, because
it is not what one would conclude were one to consider only the
dipole moment $D$, or the quadrupole moment 
$q = 1/3 - \langle \sin^2 b\rangle$.  The moments for the 1024~ms
data we are considering are $D=0.0065 \pm 0.027$ and 
$q = -0.025\pm 0.013$ (the quoted errors are statistical only;
additional error due to burst direction uncertainty is negligible).
The predicted moments for isotropic models (i.e., the moments of
the sky exposure) are $D=-0.0127$ and $q=-0.0048$.  The predicted
moments for the best-fit two component model are
$D=0.022$ and $q=-0.0024$; each is further from the data than
are the values predicted by isotropic models.  Yet the anisotropic
model has a larger likelihood than its isotropized version,
presumably because the likelihood uses much more angular information
than the dipole or quadrupole moments, including how the anisotropy
varies with flux.

Figure~14 shows contours of the joint distribution for the halo
fraction, $f$, and the cosmological
luminosity, $\nu_c$, conditional on $\nu_h = 0.01$.  
Very large halo fractions, of order 40 to 50\%, are favored.
Figure~15 shows the observed and predicted cumulative peak
flux distributions, and Figure~16 shows the differential rates
for the halo (dashed) and cosmological (dotted) components, as
well as the total differential rate (solid).
In contrast to models with luminous halo bursts, here the halo
component is most important at {\it low} fluxes.  Where the
cosmological component flattens substantially at low fluxes,
the $\F^{-5/2}$ halo component becomes important; the sum of
the two components falls with
roughly the $\F^{-2}$ behavior needed to fit the
low-flux part of the intensity distribution.  At large fluxes
both components fall like $\F^{-5/2}$, the cosmological
component dominating.

\subsubsection{Intermediate Case}

As noted above and displayed in Figure~10, the profile likelihood
for $\nu_h$ based on the 1024~ms data exhibits a small mode at
$\nu_h=1.49$; for completeness we describe the properties of these
solutions.  The best-fit parameter values are $\nu_c = 0.46$,
$\nu_h=1.49$ (implying $r_{\rm fid}=10.4$~kpc), and $f=0.05$.  
Figure~17 shows contours of credible
regions for $f$ and $\nu_c$, presuming $\nu_h$ equals its best-fit
value.  It has a shape and location intermediate to that exhibited
by the dim and luminous models discussed above.  Figure~18a compares
the observed cumulative intensity distribution with that predicted
by the best-fit model.  Figure~18b shows the local, cosmological, and
total differential rates.  As is clear from these Figures, the
local population plays a very minor role in these models.  Interestingly,
when we artificially isotropize this model, the likelihood
{\it decreases} by a factor of 2.6.  The anisotropy of the local
component helps these models, though not significantly; in this
respect they resemble the models with a dim local component.


\section{Summary and Discussion}

Our analysis of halo models in \S~2 demonstrates that such models can
account for the 3B data as successfully as cosmological models, provided
one considers halos with core sizes significantly larger than those
used to model the distribution of dark matter.  Only a few years ago,
a local population of sources with such a large characteristic size
would have been considered highly implausible; such a population would
have to have been
hypothesized purely for the purpose of hosting bursts.  We now know
that there is a population of high velocity neutron stars that could
conceivably provide a host population with a very large characteristic
length scale.  Whether the characteristics of such a population could
model the burst data as successfully as the {\it ad hoc} halo models
considered here is an open question, requiring detailed modelling
beyond the scope of this study.  Our work demonstrates how the analysis
of such models can best be undertaken, and the results of \S~2 should
guide the study of other, more complicated local models.

The core sizes we infer are smaller than one might expect
based on popular semiquantitative arguments that consider the superposed
dipole moments of shells centered on the Galactic center (Hartmann
et al.\ 1994; Briggs et al.\ 1994).  Such arguments are misleading,
as we show in \S~2.  
They fail to distinguish the artificial anisotropy arising from
the displacement of the centers of Galactocentric shells from the observing 
point (the Sun), and the actual, intrinsic anisotropy best quantified by 
calculating angular moments of shells at a constant radius from the Sun.

Our analysis of halo models also demonstrates that the 3B data do
not constrain the width of power-law luminosity functions for burst
sources.  This result contradicts the findings of Hakkila et al.\ (1995),
who used less rigorous analysis methods and who restricted
their search of parameter space to a significantly smaller region than
that explored here.  We elucidate the qualitative reasons
for the lack of such a constraint in \S~2.

As with the isotropic models studied in Paper~II, inferences based on
the 64~ms and 1024~ms data are formally inconsistent, in the sense that
there is negligible overlap of the credible regions found by analyzing
the two data sets.  Although the shapes of the burst distributions that
best model each data set differ somewhat, the inconsistency arises
largely because the two data sets imply very different burst rates
per unit volume.  The 64~ms data implies rates about 40\% larger than
the 1024~ms data, and in this sense the 64~ms timescale is {\it more} 
sensitive than the 1024~ms timescale, even though the 1024~ms data
set is larger (i.e., the 1024~ms data set is not as large as one would
expect from extrapolating the 64~ms data set to the lower fluxes
detectable using the 1024~ms timescale).
In Paper~II we argue that this difference could
arise from ``peak dilution'' in the 1024~ms data set:  when bursts
have peaks briefer than 1024~ms, their intensities are underestimated
by 1024~ms measurements.  This changes the shape and normalization
of the intensity distribution in a manner that may account for the
discrepancy between the two data sets.  The 64~ms data are not immune
to such effects, but will be less affected.

We also studied two-population models, consisting of superposed 
standard candle cosmological and local halo populations.  
Numerous studies
suggest that there may be two (or more) classes of bursts, as we
review in \S~3.  These models also
serve a purely pragmatic purpose of allowing precise, model-dependent
quantification of the constraints the 3B data place on the anisotropy
of the distribution of burst sources.  They also provide a
model-dependent test of the adequacy of cosmological models.
For the two-population
models, we took the halo population to follow the distribution of dark matter
in a standard Bahcall-Soneira halo with a core size of 2~kpc, so
that the local population is associated with a known distribution
of matter.  Two families of models successfully account for the
data:  models with luminous halo sources visible to $\sim 50$~kpc;
and models with dim halo sources visible from within a disk scale
height.  Despite the fact that the luminous halo sources would be distributed 
anisotropically,
models with $\approx 10$\% of observable bursts from the halo are
favored, and halo fractions as large as 20\% are acceptable.
Dim halo sources would comprise a nearly isotropic, homogeneous component.
For such sources, the data favor large halo fractions, of the order 
of 40\% to 50\%.  
These results are consistent with the relative sizes 
of classes of bursts inferred from characteristics of burst lightcurves,
or with the fraction of bursts observed by {\it Ginga} to have low energy
absorption features.  We have not yet ascertained whether membership
in these classes is correlated with burst intensity in the manner
that these models would predict.

The common methodology employed here and in Paper~II, where we analyze
cosmological models, permits us to rigorously compare how well cosmological
and local models account for the full joint distribution of burst
peak fluxes and directions.  The Bayesian tool for this comparison---the
Bayes factor---objectively accounts for parameter uncertainty, providing
a quantitative ``Ockham's Razor.''  We find that the data do not
decisively prefer any one of the models we have studied to its competitors.
In particular, local models and models with a substantial local
component account for the 3B burst intensity and direction data 
as well as purely cosmological models. 

Despite this, several investigators strongly prefer cosmological models
to local ones, or vice versa.  This preference can be justified only by
consideration of information beyond that in the distribution of burst
peak fluxes and directions analyzed here.  Numerous studies have been
undertaken of some additional burst characteristics, including searches
for evidence of time dilation, burst repetition, and spectral lines.
To date, such searches have been inconclusive, with investigators who
use different frequentist methodologies arriving at different, often
conflicting conclusions.
The Bayesian approach we have adopted in this work
can be straightforwardly generalized to address many of these issues.
Some such generalizations have already been outlined in Paper~I,
and we are pursuing studies of some of these outstanding controversial
issues from within this framework.

\acknowledgments

This work was supported in part by NASA grants
NAG 5-1758, NAG 5-2762, NAG 5-3097, and NAG 5-3427; 
and by NSF grants AST91-19475 and AST-93-15375.


\appendix
\section{Constraining Model Parameters}

In this
work we have addressed questions about the adequacy of
parameterized models.
Statisticians using both frequentist and Bayesian methods have divided 
such questions into two classes.
First is the class of {\it estimation} questions that assess the implications
of assuming the truth of a particular model, usually by estimating values
or allowed ranges for the model parameters.  Second is the class of
{\it model assessment} questions that assess the viability of a model.
We have outlined Bayesian methods for treating these questions in
Paper I and Paper II.  A clear discussion of the application of frequentist 
methods for estimation and model
assessment to problems in the physical sciences
is available in the text by Eadie, et al.\ (1971).

The procedures used for estimation are fundamentally different from those
used for model assessment.  Unfortunately, the complexity of the GRB
data makes the distinction between these problems somewhat subtle from
the frequentist viewpoint, enough so 
that nearly every previously published statistical analysis of these data has
inappropriately used model assessment procedures to address estimation
problems.  In particular,
a number of studies used goodness-of-fit (GOF) procedures to specify
``confidence'' regions, usually based on the $\chi^2$ or 
Kolmogorov-Smirnov (KS) GOF statistics.
In these studies, the boundary of the calculated ``confidence region'' was
determined by finding parameters for which the significance level of
a GOF test is equal to the desired confidence level.  Such misapplication
of GOF procedures to parameter estimation problems is commonplace in
astrophysics; we have been guilty of it ourselves in the past.
Thus some discussion of the problems with this approach
seems worthwhile.

To see the problems
with such a procedure, it is instructive to apply this type of thinking
in a context where the correct procedure is widely known.
Consider, therefore, estimation and model assessment for a model
with a single parameter, $\mu$, using the $\chi^2$ statistic.
A common best-fit estimate for $\mu$ is the minimum $\chi^2$ value,
which we will denote by $\hat\mu$.  A confidence region (CR) for $\mu$ is
typically specified by reporting the region of $\mu$ for which
$\chi^2$ is less than some critical value, $\chi^2_{{\rm CR},P}$,
found by adding to $\chi^2_{\rm min} = \chi^2(\hat\mu)$ a fixed
number, $\Delta\chi^2_P$, whose value is determined by the size of
the confidence region desired:  
\beq
\chi^2_{{\rm CR},P}(\hat\mu) = \chi^2(\hat\mu) + \Delta\chi^2_P.
  \eqn(chi2-CR)
\eeq
For example, for a 68\% confidence
region, $\Delta\chi^2_P=1$ (asymptotically).  Note that the value
of the critical $\chi^2$ defining the region depends on the data
(through $\hat \mu$).

Consider now the different problem of assessing the goodness-of-fit
of some hypothesis that specifies a particular value of $\mu$
a priori; we denote this value by $\mu_0$.
For this problem, one typically finds the critical value,
$\chi^2_{{\rm GOF},P}$, such that one expects $\chi^2(\mu_0)$ to be less
this value $100 P$\% of the time, presuming all model assumptions are
true.  By agreeing to reject the model if $\chi^2(\mu_0)>\chi^2_{{\rm GOF},P}$,
one will falsely reject a true model (i.e., commit a ``Type I error'')
a fraction $(1-P)$ of the time.  
In contrast to the CR case, the critical value defining a GOF
test is a constant that depends only
on the number of the data, and not on the actual data values.
It is simply the $\chi^2$ value such that the probability for
observing larger values from the $\chi^2_\nu$ distribution with
$\nu=N$ is $(1-P)$; for example, for $P=0.95$, 
$\chi^2_{{\rm GOF},P} = 43.8$ for $N=30$.
(If instead of an a priori $\mu$ value we wish to test the
adequacy of the best fit model, we would compare $\chi^2_{\rm min}$
with a critical value from the $\chi^2_\nu$ distribution with $\nu=N-1$.)

Previous analyses of halo models used methods that correspond to
finding a $P=0.68$ confidence region for $\mu$ by reporting a region
bounded by values of $\mu$ such that $\chi^2(\mu) = \chi^2_{{\rm GOF},P}$,
rather than $\chi^2(\mu) = \chi^2_{\rm min} + 1$.  One problem with
such a procedure is immediately apparent:  
for nearly a third of all data sets, no such ``confidence region'' will
exist because for 32\% of all Gaussian data, $\chi^2_{\rm min}$ will
be greater than the critical value.  In fact, for every data set, there
will be some constant, $C$, such that no ``confidence region''
exists for any confidence level less than $C$.  For those data sets
for which such a ``confidence region'' exists, it will vary in size,
sometimes being larger than the correct region, and
sometimes smaller.

One might hope that there might be some relationship between the regions
such that, on the average, they might coincide.  To see that this hope
is forlorn, consider a specific, simple problem whose solution is obvious.
The simple problem we will discuss
is inference of some quantity of unknown magnitude, $\mu$, from
data consisting of $N$ measurements of $\mu$, each contaminated
with added noise.  We will denote the measured values by $x_i$,
and model them as being the sum of $\mu$ and noise components
to which we assign independent Gaussian probabilities with zero
mean and known, common standard deviation, $\sigma$.
For this problem, both a frequentist analysis and a Bayesian analysis
with a broad prior lead to a best estimate of $\mu = \bar x$ and
a 68\% confidence or credible region of 
$\bar x \pm \sigma/\surd N$, where $\bar x$ is the sample mean.
We can connect this problem to the above discussion by examining the
likelihood, which is simply the product of
$N$ Gaussian distributions,
\beq
\like(\mu) = {1 \over \sigma^N (2\pi)^{N/2}} 
\exp\left[-{\sum (x_i - \mu)^2 \over 2\sigma^2}\right].\eqn(g-like1)
\eeq
Its implications for $\mu$ can be more clearly seen by expanding 
the square in the exponential and then completing the square
in $\mu$, which gives,
\beq
\like(\mu) = {1 \over \sigma^N (2\pi)^{N/2}} 
\exp\left[-{N s^2 \over 2\sigma^2}\right]
\exp\left[-{N(\bar x - \mu)^2 \over 2\sigma^2}\right],\eqn(g-like)
\eeq
where $s^2 \equiv \sum (x_i-\bar x)^2/N$ is the sample variance.  
As a function of $\mu$, this is proportional to a Gaussian centered
at $\bar x$ with a width of $\sigma/\surd N$.
Its relationship to the $\chi^2$ procedure just described is obvious
once we note that, up to a term
constant in $\mu$, the log likelihood is proportional to $\chi^2$:
\beqa
-2\log \like(\mu) &=& \sum {(x_i-\mu)^2 \over \sigma^2} + C\\
&=& \chi^2 + C.\eq(L-chi)
\eeqa
Using these results,
it is easy to verify that the $\chi^2_{\rm min}+1$ interval is
exactly the familiar $\bar x \pm \sigma/\surd N$ interval.  More
importantly, for this simple example we can examine the behavior
of the incorrect procedure, where the ``confidence region'' is
bounded by the 68\% critical $\chi^2$ value.  The half-width of
the incorrect region can be found analytically; it is given by
\beq
\delta = {\sigma \over \sqrt{N}} 
   \left[\chi^2_{68} - \chi^2_{\rm min}\right]^{1/2}.\eqn(bad-cr)
\eeq
A good estimate of the average behavior of the region can be found
by substituting the expected value of $\chi^2_{\rm min}$ (equal
to $N-1$) and an asymptotic expression for $\chi^2_{68}$
(equal to $N+0.47\sqrt{2N}$), giving
\beq
\langle\delta\rangle \approx {\sigma \over \sqrt{N}}
   \left[1 + 0.47\sqrt{2N}\right]^{1/2}\eqn(exp-delta)
\eeq
(we have verified the accuracy of this expression with Monte Carlo
simulations).  This gives $\langle\delta\rangle \approx 1.8 \sigma/\surd N$
for $N=10$, and $\langle\delta\rangle \approx 2.8 \sigma/\surd N$
for $N=100$.  It is thus clear that, on the average, the region will be 
significantly too large, and that the discrepancy actually grows
with the number of data.  Finally, the 68\% probability associated with
this region is obviously not the covering probability for this procedure,
since we know the smaller $\sigma/\surd N$ region has this covering
probability.  These results clearly 
show that the use of GOF procedures
to find ``confidence regions,'' however intuitively appealing it may at first 
seem, is incorrect.

It is interesting to speculate about why such a basic mistake is so frequently
made.  One reason is that, for the familiar
case of Gaussian statistics, the same function---$\chi^2$---is used
both to define the point-valued statistic used in a GOF test
($\chi^2_{\rm min}$ or $\chi^2(\mu_0)$), and the interval-valued
statistic used for a confidence region (the $\mu$ range with
$\chi^2<\chi^2_{\rm min}+\Delta\chi^2$).  This may have been why some
investigators used the KS GOF statistic to constrain parameters,
although we know of no statistical literature suggesting that
this statistic is useful for estimation problems.  More fundamentally,
the confusion may arise because there are several qualitatively different
probabilities in frequentist statistics.  Covering probabilities for
confidence regions, Type I error probabilities, Type II error 
probabilities---all of these are quantities that span $[0,1]$ that
scientists can use to assess the reasonableness of hypotheses.  But none
of them are probabilities {\it for hypotheses}, so one may easily
be confused about which is most closely related to the question
one is asking.  This confusion is exacerbated by the fact that all
frequentist probabilities must condition on a particular point hypothesis,
even those that refer to an entire class of hypotheses.
For some problems (particularly for confidence region calculations),
the hope is that the final result is independent of the particular
hypothesis used.  But this is seldom true in real problems, so that
one hypothesis must inevitably be chosen to represent a class
of hypotheses (e.g., approximate confidence regions are found using 
calculations conditioning on the best-fit hypothesis).

This confusion cannot arise in the Bayesian approach.  One always
calculates probabilities for hypotheses, so there is never ambiguity
over what kind of hypothesis one's probability is associated with:
one must explicitly state it in order even to start the
calculation.  If one seeks a measure of how plausible it is for a 
parameter to lie in some region, one simply
calculates the probability that it is
in that region (parameter estimation).  If instead one wishes to
assess an entire model, one calculates the probability for that
model as a whole (model
comparison).  The formalism forces one to distinguish between
these options.

\section{Properties of Halo Models With Luminosity Functions}

In this Appendix we derive the properties of the differential burst
rate for a halo population with a power-law luminosity function
mentioned in \S~2.  We focus on the slope of the intensity distribution.
Consider, therefore, the differential burst rate along a particular
line of site, $\drxn$.  Let $r$ denote the distance from the observer
along the line of site, and write the burst rate density as
$\dot n(r,\lum) = \dot n(r) f(\lum)$, where $f(\lum)$ is the
normalized luminosity function.  Then the differential rate is given by
\beq
{dR \over d\F d\drxn} 
  = \int dr\, r^2 \dot n(r) \int d\lum\, f(\lum) 
   \delta\left(\F - {\lum \over 4\pi r^2}\right).\eqn(plh-dR1)
\eeq
Changing variables in the $\delta$-function from $\F$ to $r$ and
integrating over $r$ gives
\beqa
{dR \over d\F d\drxn} 
  &=& {1 \over 2\F} \int d\lum\, f(\lum) \int dr\, r^3 \dot n(r)  
   \delta\left[r - \left(\lum \over 4\pi \F\right)^{1/2}\right]\\
  &=& {\F^{-5/2} \over 16 \pi^{1/2}}
     \int d\lum\, \lum^{3/2}\, f(\lum)\, 
     \dot n\left(\sqrt{\lum \over 4\pi \F}\right).    \eq(plh-dR)
\eeqa
As a simplified description of the halo models considered in the body
of the paper, let
\beq
\dot n(r) = \cases{\dot n_0, &if $r \le r_c$;\cr
             \dot n_0\left(r \over r_c\right)^{-2}, & if $r>r_c$.\cr}
   \eqn(ndot-simp)
\eeq
Then $dR/d\F$ is proportional to $\F^{-5/2} I(\F)$, where
\beq
I(\F) \equiv \int_{\lum_l}^{\lum_u} d\lum\, \lum^{3/2-p} \times
  \cases{1, &for $\lum \le 4\pi r_c^2 \F$;\cr
         \left(\lum\over 4\pi r_c^2\F\right)^{-1},
                 &for $\lum > 4\pi r_c^2 \F$.\cr}   \eqn(I-def)
\eeq
If $\F$ is such that the upper limit of the integral is 
less than $4\pi r_c^2 \F$, then only the first case of the
integrand is relevant and $I(\F)$ is simply a constant.
Thus so long as $\F \ge \F_b$, where $\F_b =\lum_u/4\pi r_c^2$, we find that
$dR/d\F\propto \F^{-5/2}$, the power law expected for observing
a homogeneous population.

If, on the other hand, the lower limit is greater than $4\pi r_c^2\F$, 
only the second case of the integrand is relevant, and $I(\F)\propto \F^{-1}$.
Thus $dR/d\F\propto \F^{-3/2}$ when $\F < \F_a$, where
$\F_a = \lum_l/4\pi r_c^2$.

For intermediate values of $\F$, the integrand must be broken into
separate cases:  we first integrate from $\lum_l$ to $4\pi r_c^2\F$, and
then from $4\pi r_c^2\F$ to $\lum_u$.  The result is
\beqa
I(\F)
  &=& \F^{5/2-p} \, (4\pi r_c^2)^{5/2-p}\cr
  &\quad& \times \bigg\{
     {1 \over 5/2-p}\left[1 - 
          \left(\F_a \over \F\right)^{5/2-p}\right]
     + {1 \over p-3/2}\left[1 - 
          \left(\F \over \F_b\right)^{p-3/2}\right]
      \bigg\}.  \eqn(I-result)
\eeqa
Now consider the behavior of the terms in braces for the case when
$3/2 < p < 5/2$, which is the parameter regime most relevant to
GRBs.  As long as $\F\gg\F_a$ and $\F\ll \F_b$, the $\F$-dependent
terms are negligible, so $I(\F) \propto \F^{5/2-p}$.  This implies
that $dR/d\F \propto \F^{-p}$ in this region; the $\F$-dependent
terms merely smooth the transitions from the $\F^{-3/2}$ behavior
at low fluxes, and to the $\F^{-5/2}$ behavior at high fluxes.  Figure 4
portrays these results.

\section{Dipole Moments of Halo Models}

The intrinsic dipole moment of a shell observed at a radius $r'$ from
the Sun is
\beq
D(r') = { \int_{-1}^{1} d\mu \, (r')^2 \mu \dot n'(r',\mu) \over
          \int_{-1}^{1} d\mu \, (r')^2  \dot n'(r',\mu)},\eqn(D-def)
\eeq
where $\dot n'(r,\mu)$ is the burst rate per unit volume at a distance
$r'$ along any direction that makes an angle $\theta = \cos^{-1}\mu$ with
the Galactic center.  We presume here a burst rate that is spherically
symmetric about the Galactic center, so no azimuthal dependence enters.
If the burst rate per unit volume at a Galactocentric distance $r$
is $\dot n(r)$, then by the law of cosines
\beq
\dot n'(r',\mu) = \dot n[(r_0^2 + (r')^2 + 2r'r_0\mu)^{1/2}].\eqn(np-def)
\eeq
For the halos we consider above, $\dot n(r) \propto 1/(r^2 + r_c^2)$,
and the integrals required to evaluate $D(r')$ can be done analytically,
giving
\beq
D(r') = {1 \over 2}\left[\rho' + {1\over\rho'} + {\rho_c^2\over \rho'}
  - {4 \over \ln\left( (\rho'+1)^2 + \rho_c^2 \over (\rho'-1)^2 + \rho_c^2\right)}  \right],\eqn(D-exact)
\eeq
where $\rho' = r'/r_0$ and $\rho_c = r_c/r_0$.
This expression is somewhat complicated, and an amazing amount of
cancellation occurs between the various terms.  We gain more insight into the
behavior of $D(r')$ by approximating it in the limit where $r_0$ is
small compared with the other length scales ($r'$ and $r_c$).  Due to
the high-order cancellation among the terms in equation~\ceq{D-exact},
the approximate result is best found by separately approximating the integrals
in equation~\ceq{D-def}.  The result is
\beq
D(r') \approx {2 \over 3}\, {1 \over \rho' + {1\over \rho'} + {\rho_c\over \rho'}},\eqn(D-approx)
\eeq
as cited in the main text.  The approximation is surprisingly good for
all ranges of the parameters discussed in this work.


\clearpage

\figcaption{Credible regions for the dimensionless luminosity, $\nu_h$, and 
dimensionless core size, $\rho_c$,
of a halo model, based
on the 64~ms (a) and 1024~ms (b) data.  Throughout
this work crosses show the best-fit
parameter values; contours bound the 68.3\% (dotted), 95.4\% (dashed),
and 99.7\% (solid) credible regions of highest posterior density.}

\figcaption{As in Fig.~1, but for models omitting the halo around M31.}

\figcaption{As in Fig.~1, but for models using a $1/(1+r/r_c)$ halo.
Only results from the 1024~ms data are shown.}

\figcaption{Joint credible regions for the burst rate density $\dot n_0$
and luminosity parameter $\nu_h$, conditional on the best-fit values
of the core size for each data set (a), and on a common value
of $\rho_c = 10$ (b).} 

\figcaption{Schematic behavior of the differential burst rate
for a halo with a bounded 
power law luminosity function proportional to $\Lambda^{-p}$.} 

\figcaption{Log-likelihood vs.\ dimensionless upper luminosity
$\nu_u$ for models with a $\Lambda^{-1.8}$
luminosity function with a dynamic range of $10^3$.  Results are
shown for both the 64~ms (solid) and 1024~ms (dashed) data.
({\it a})--- Results with $\rho_c$
fixed to the best-fit standard candle model values
($\rho_c=18.1$ for 64~ms, 7.8 for 1024~ms).  ({\it b})---Results
with $\rho_c$ slightly reduced (to 10 for
the 64~ms data, and 5 for the 1024~ms data).
}

\figcaption{Visibility of large Galactocentric shells.  Only
the highlighted part of a Galactocentric shell of
radius $r$ is visible, if $r$ is comparable to the maximum solarcentric radius,
$r'$, from which bursts are visible.  The visible portion has a
large negative dipole moment.}

\figcaption{Burst rate per unit radius, and dipole moment of 
solarcentric shells of radius $r'$ for standard candle models
(a, b) and models with a broad luminosity function (c, d).}

\figcaption{Dipole moments of the models of Figure 8, versus
peak flux rather than radius.  Points with error bars denote
the measured dipole moments of 5 subsets of the data in bins
of equal logarithmic width.}

\figcaption{Profile likelihood vs.\ the dimensionless luminosity $\nu_h$
of the halo component of two-population models, for 64~ms (solid) and
1024~ms (dashed) data.}

\figcaption{Joint credible regions for $f$ and $\nu_c$ for two-population
models with luminous halo bursters, based on the 64~ms (a) and
1024~ms (b) data.}

\figcaption{Predicted (smooth curve) and observed (histogram) 
cumulative intensity distributions
of best-fit two-population models with luminous halo bursters, for 64~ms (a)
and 1024~ms (b) data.}

\figcaption{Differential rates for the halo (dashed) and cosmological
(dotted) components comprising the models of Figure 12.  The solid
curve shows the total differential rate.}

\figcaption{As Fig.~11, but for models with dim halo bursters.}

\figcaption{As Fig.~12, but for models with dim halo bursters.}

\figcaption{Differential rates for the halo (dashed) and cosmological
(dotted) components comprising the models of Figure 12.  The solid
curve shows the total differential rate.}

\figcaption{Credible regions for $f$ and $\nu_c$ for 
two-population models whose halo component has an intermediate
luminosity ($\nu_h=1.5$, corresponding to the middle mode
in Fig.~10), base on 1024~ms data.}

\figcaption{The predicted peak flux distribution for 
the best-fit $\nu_h=1.5$ two-population model for the 1024~ms data.
(a) Predicted (smooth curve) and observed (histogram) 
cumulative intensity distributions.  (b) Total differential rate
(solid), with halo (dashed) and cosmological (dotted) components
plotted separately.}

\clearpage

\begin{deluxetable}{lrr}
\tablecolumns{3}
\tablewidth{0pc}
\tablecaption{Standard Candle Halo Models}
\tablehead{
\colhead{Quantity}    &   \colhead{64~ms Results}   &
\colhead{1024~ms Results} } 
\startdata
$\rho_c$ & 18.1 & 7.8 \nl
$\nu_h$ & $2.20\times 10^3$ & $3.50\times10^2$ \nl
$\dot n_0$ (yr$^{-1}$~kpc$^{-3})$ & $4.2\times 10^{-6}$ 
           & $3.7\times 10^{-5}$ \nl
$R_{9,4}$ & 1.11 & 5.05 \nl
$p(>R_{9,4})$ & \nodata & \nodata \nl
$B_{9,4}$ & 0.062 & 0.49 \nl
\enddata
\end{deluxetable}

\begin{deluxetable}{lrr}
\tablecolumns{3}
\tablewidth{0pc}
\tablecaption{Two-Component Models}
\tablehead{
\colhead{Quantity}    &   \colhead{64~ms Results}   &
\colhead{1024~ms Results} } 
\startdata
\cutinhead{$M_{11}$: Luminous Halo Sources} \nl
$\nu_c$ & 0.16 & 0.37 \nl
$\nu_h$ & 34.3 & 19.1 \nl
$f$ & 0.11 & 0.073 \nl
$R_{11,4}$ & 3.19 & 2.14 \nl
$p(>R_{11,4})$ & 0.31 & 0.47 \nl
$B_{11,4}$ & 0.45 & 0.25 \nl
\cutinhead{$M_{12}$: Dim Halo Sources} \nl
$\nu_c$ & 3.82 & 1.38 \nl
$\nu_h$ & $\equiv 0.01$ & $0.01$ \nl
$f$ & 0.59 & 0.36 \nl
$R_{12,4}$ & 8.25 & 46.1 \nl
$p(>R_{12,4})$ & 0.12 & 0.022 \nl
$B_{12,4}$ & 2.5 & 6.7 \nl
\enddata
\end{deluxetable}

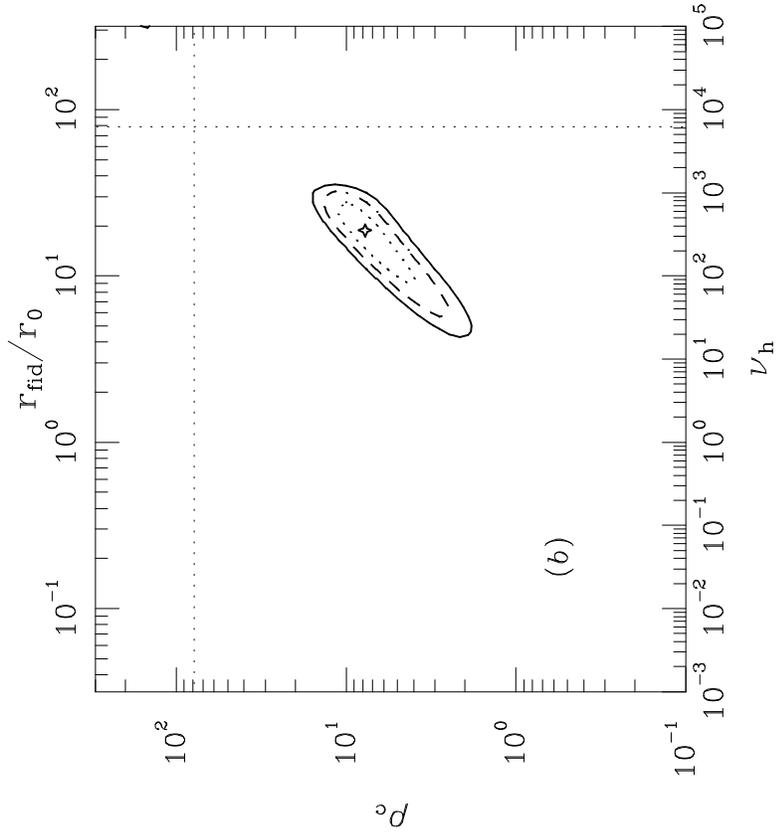
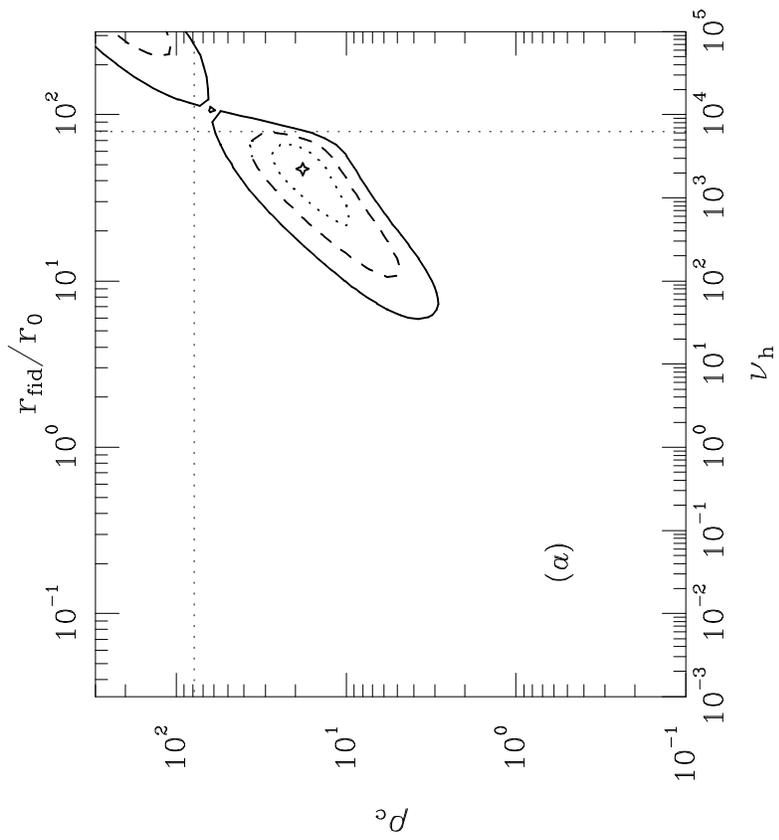
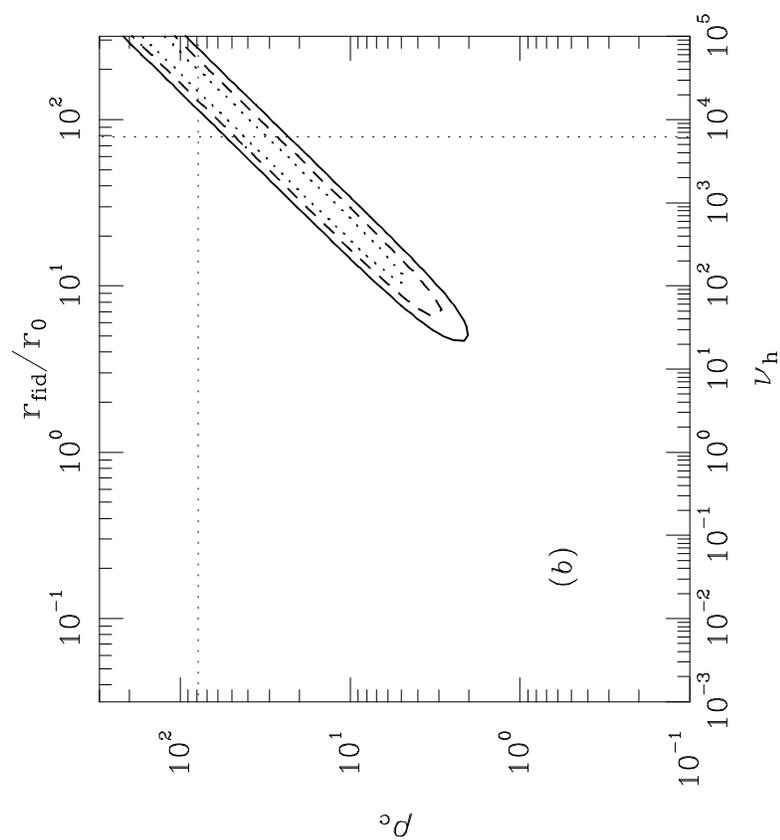
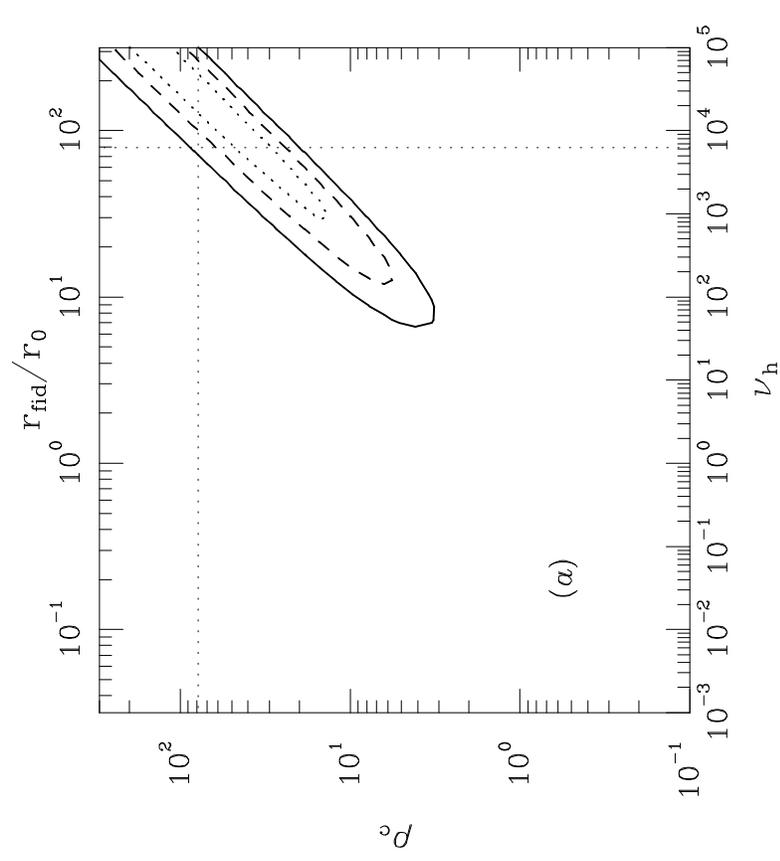
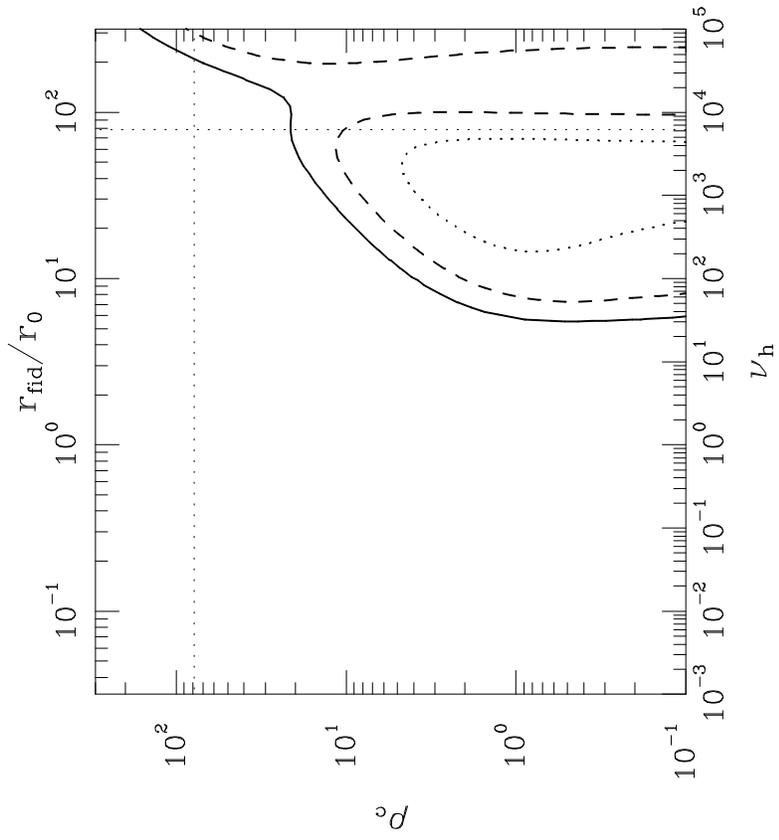
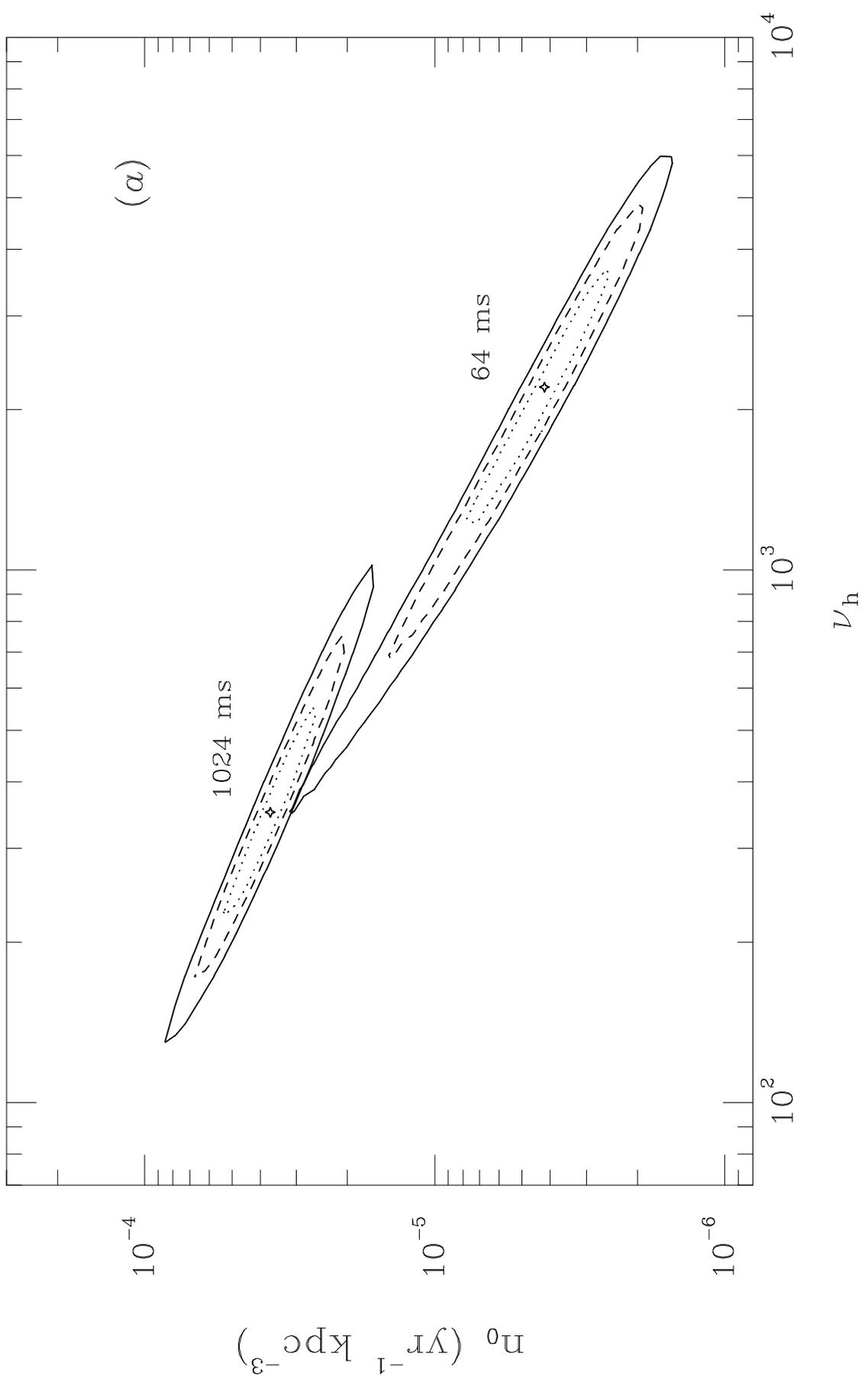
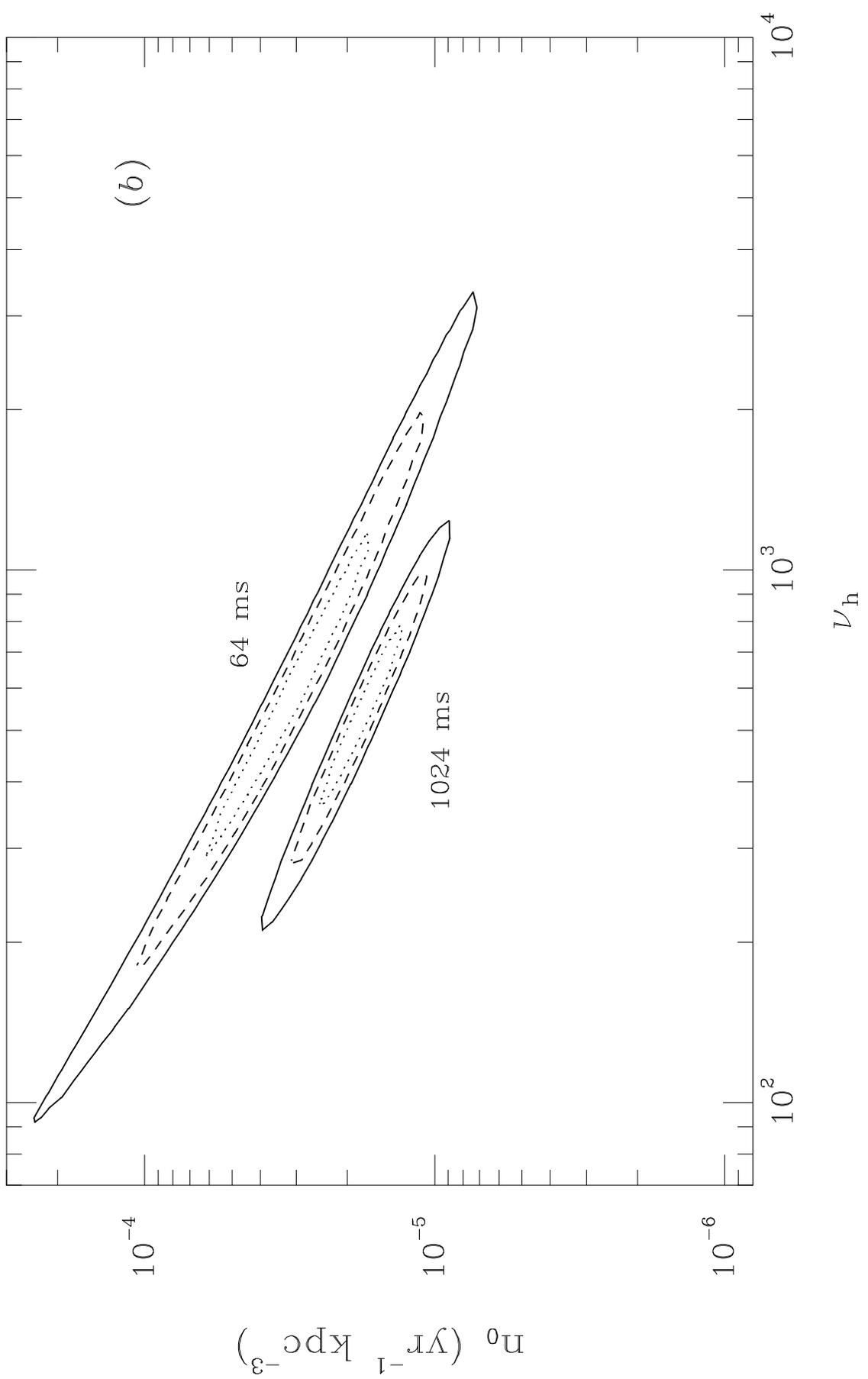
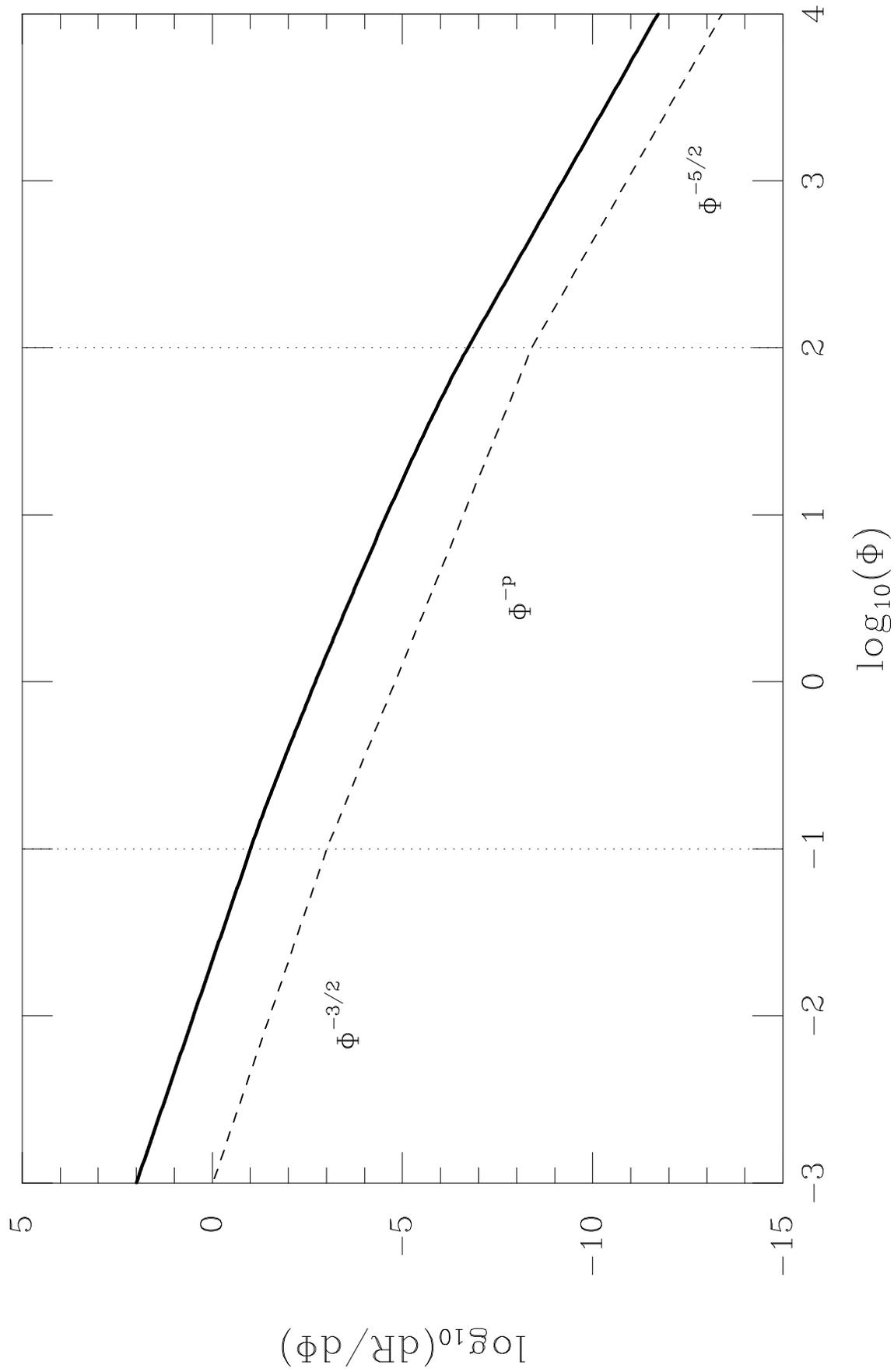
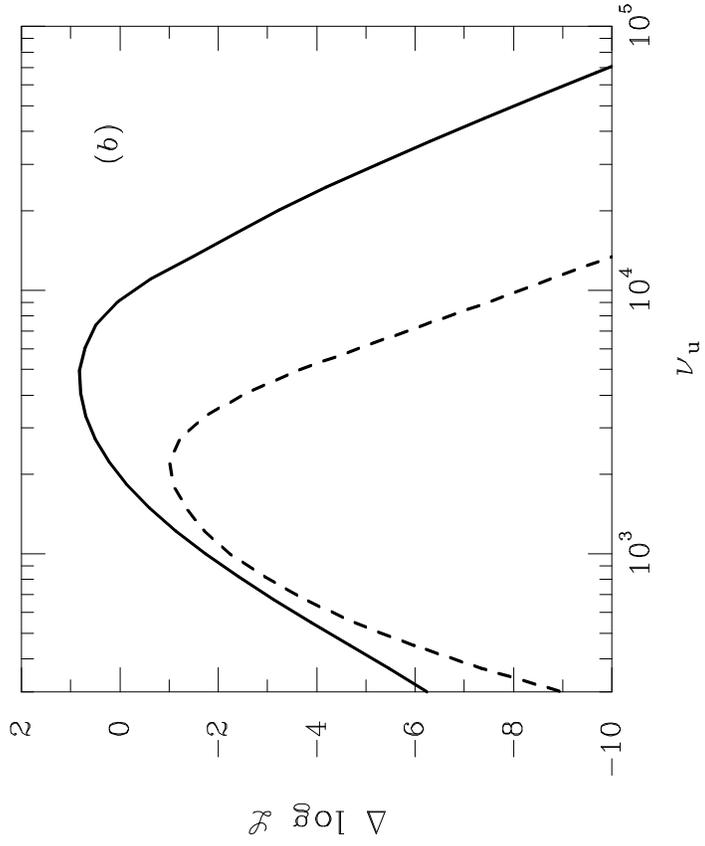
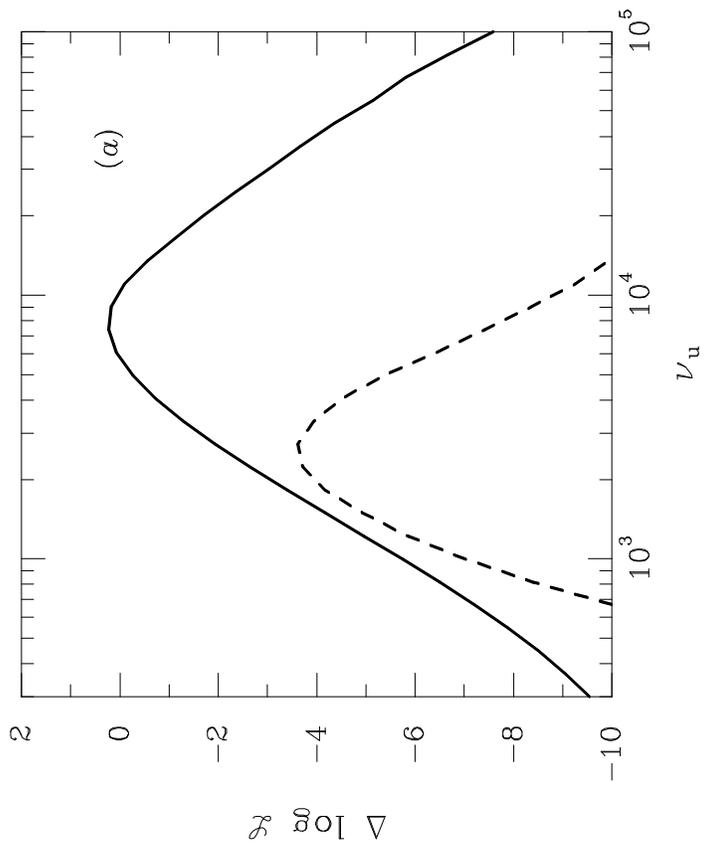
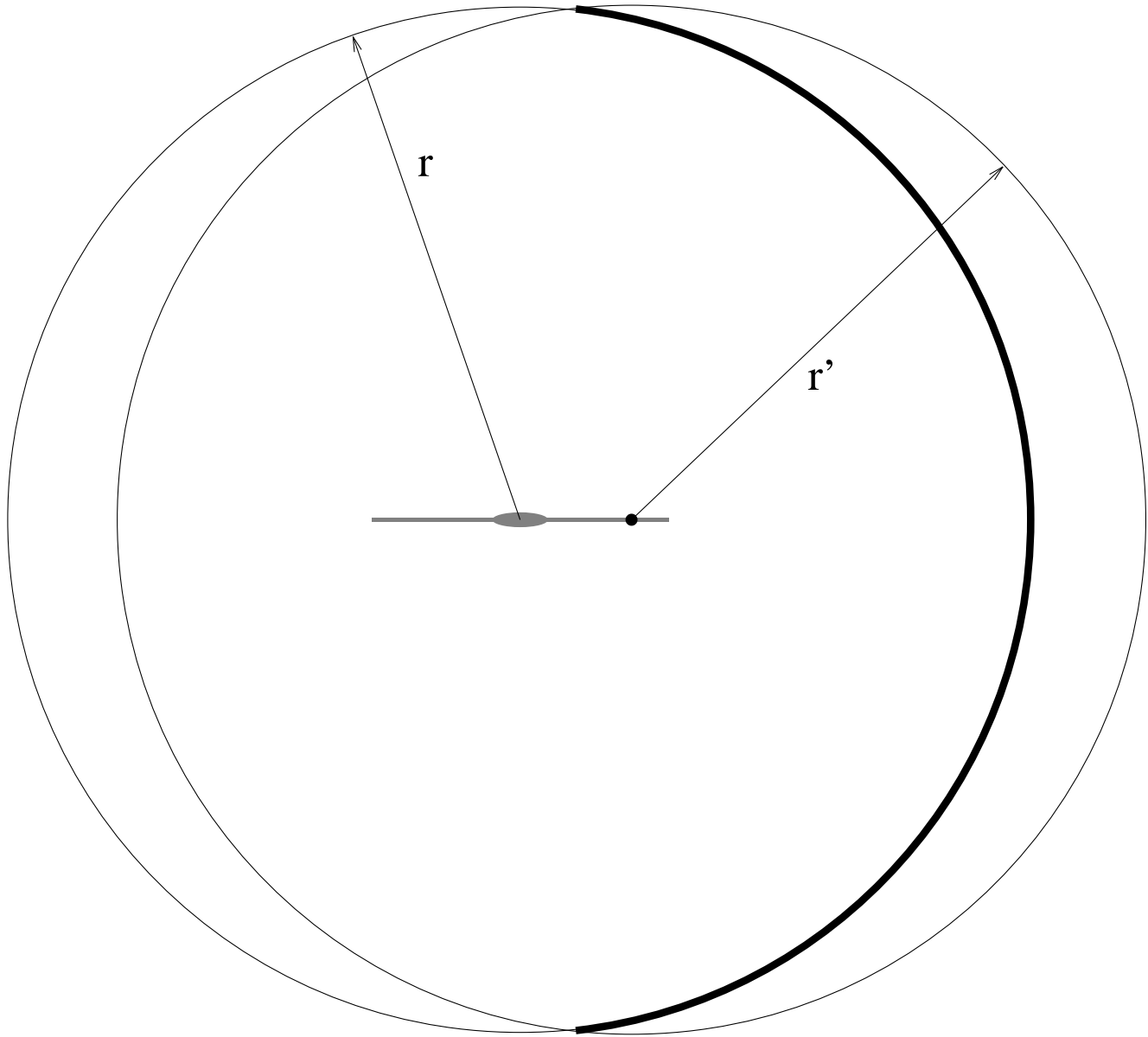
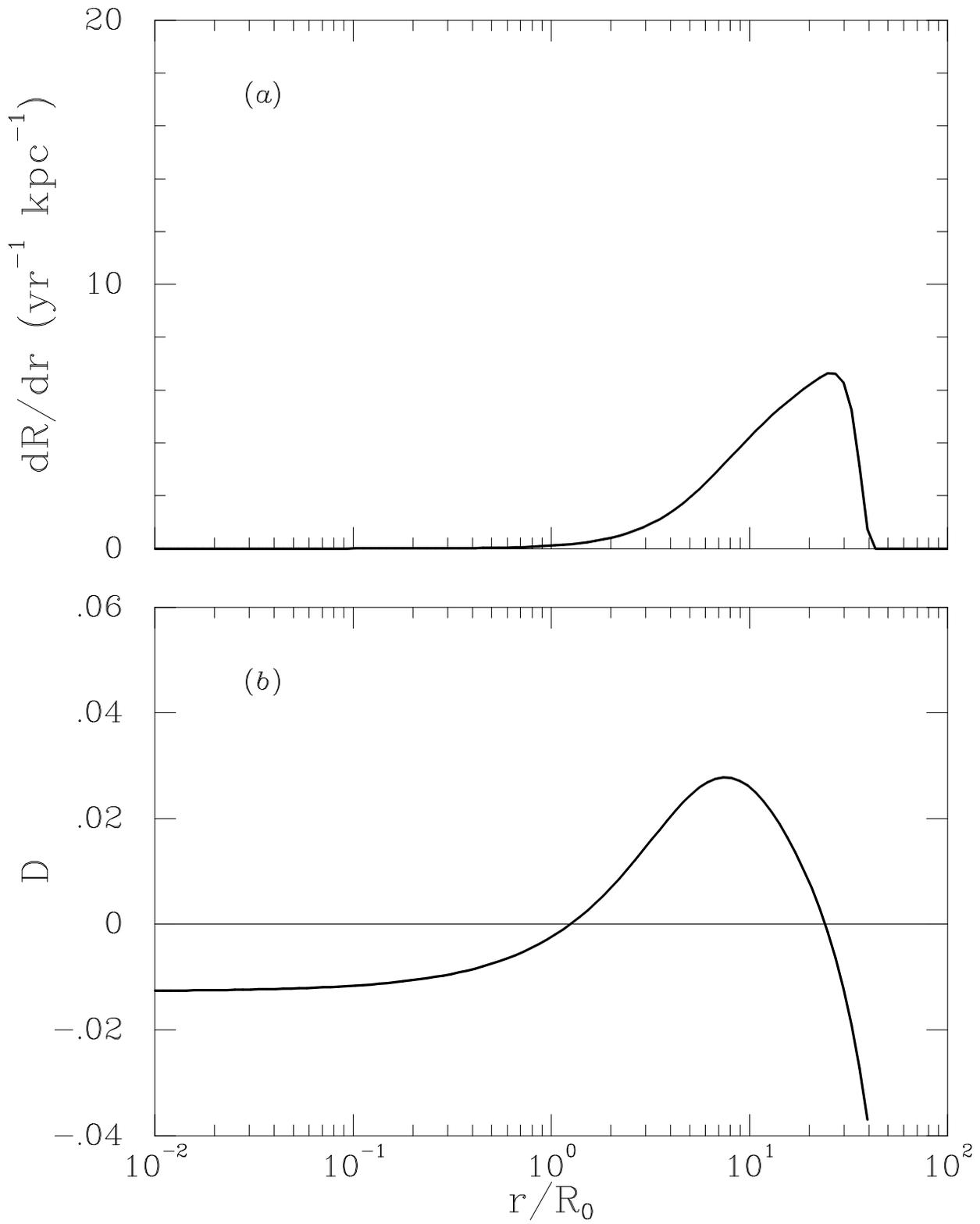
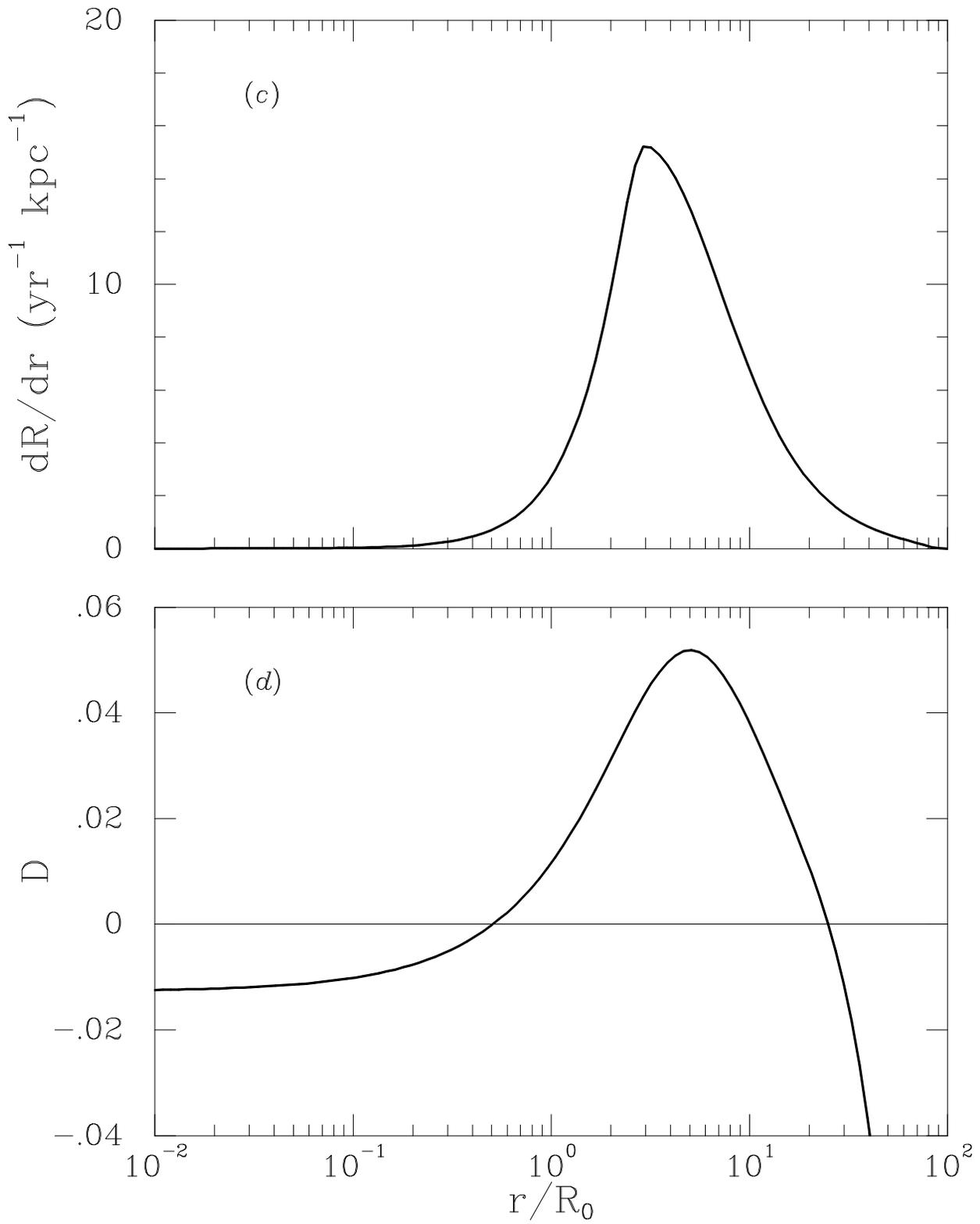
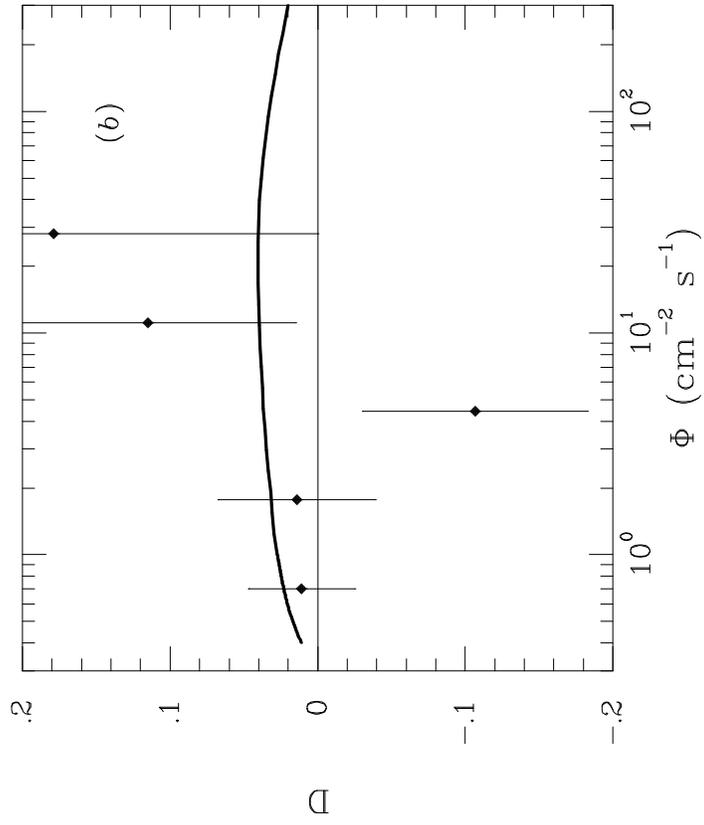
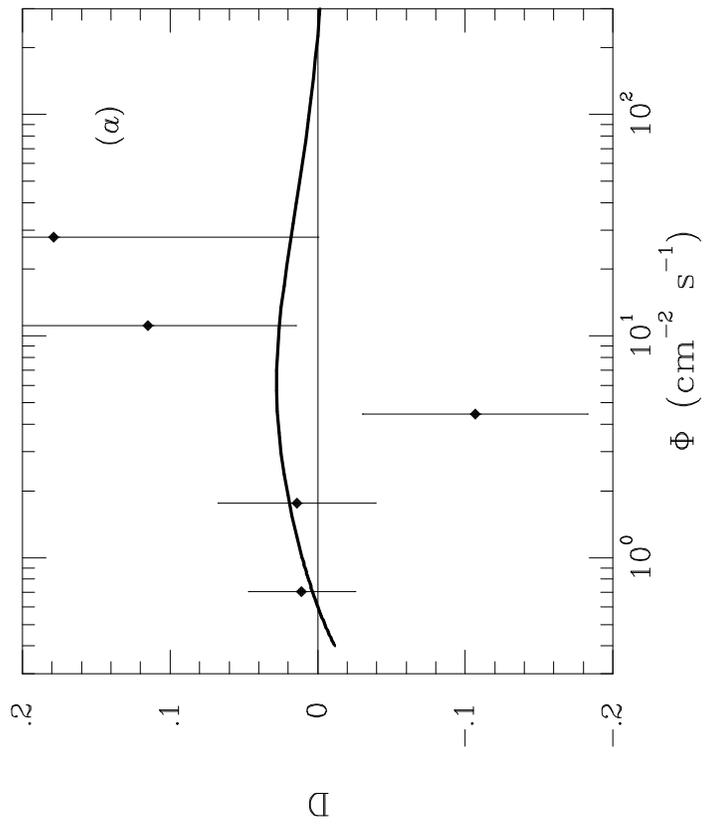
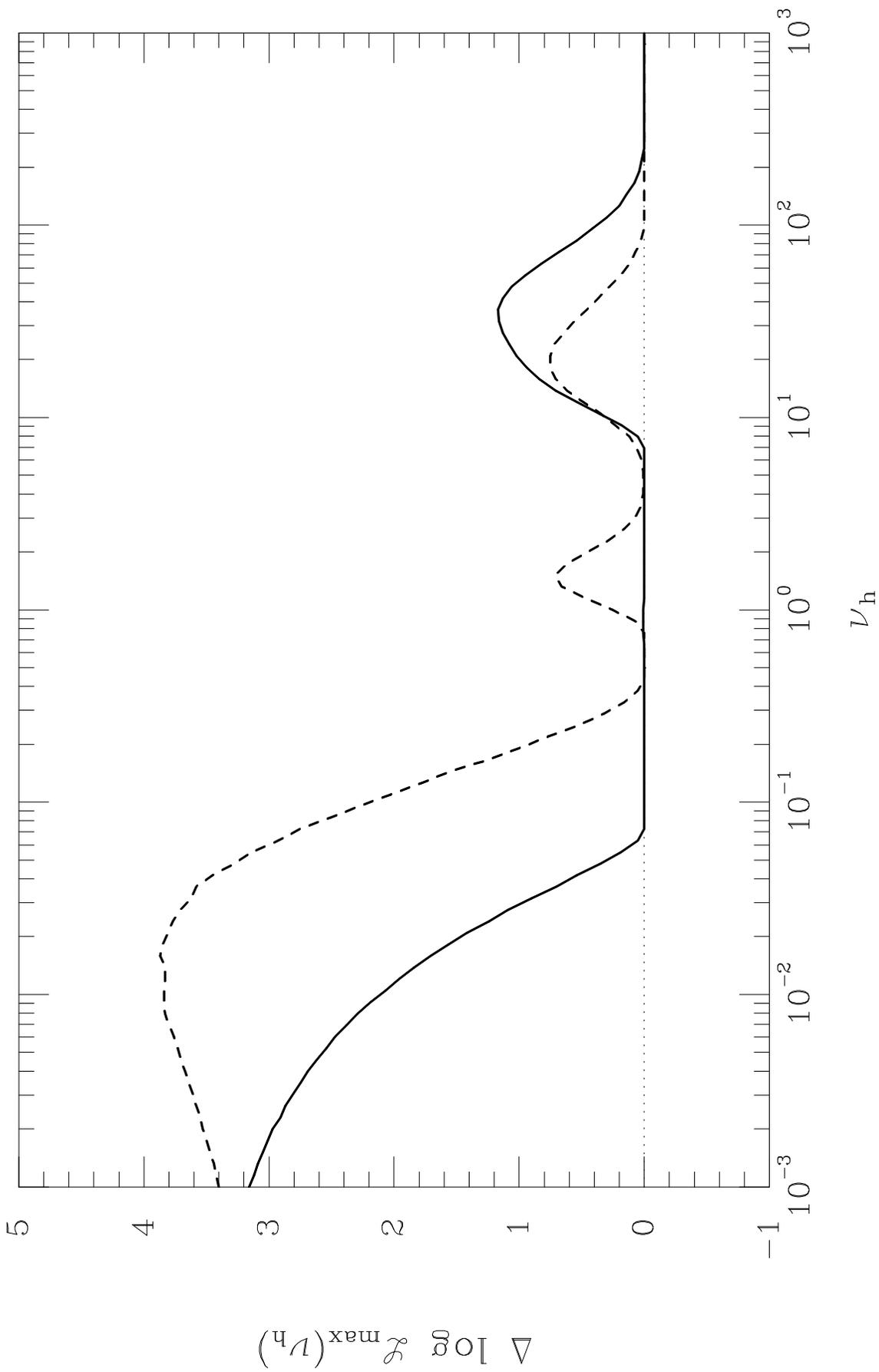
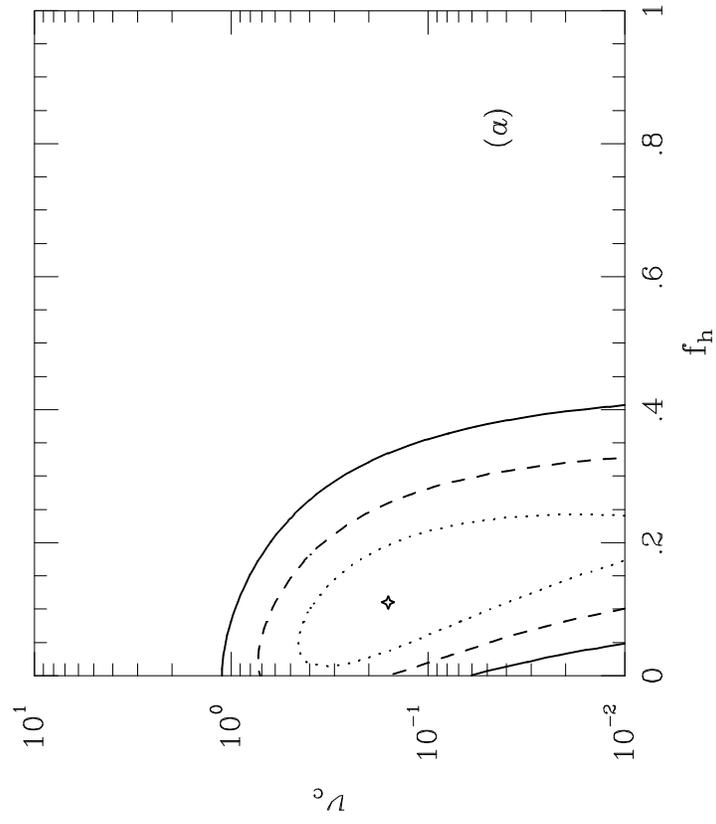
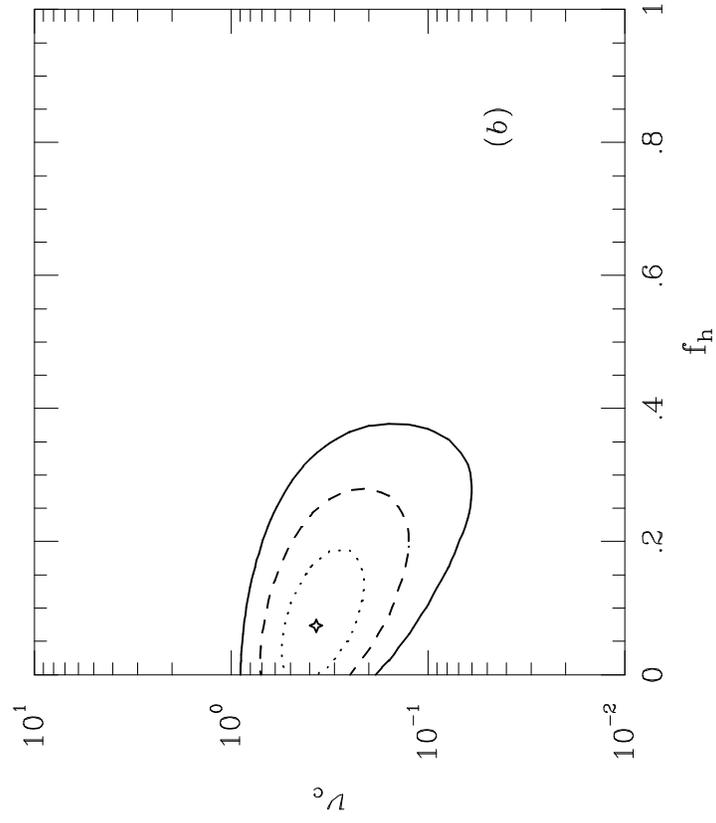
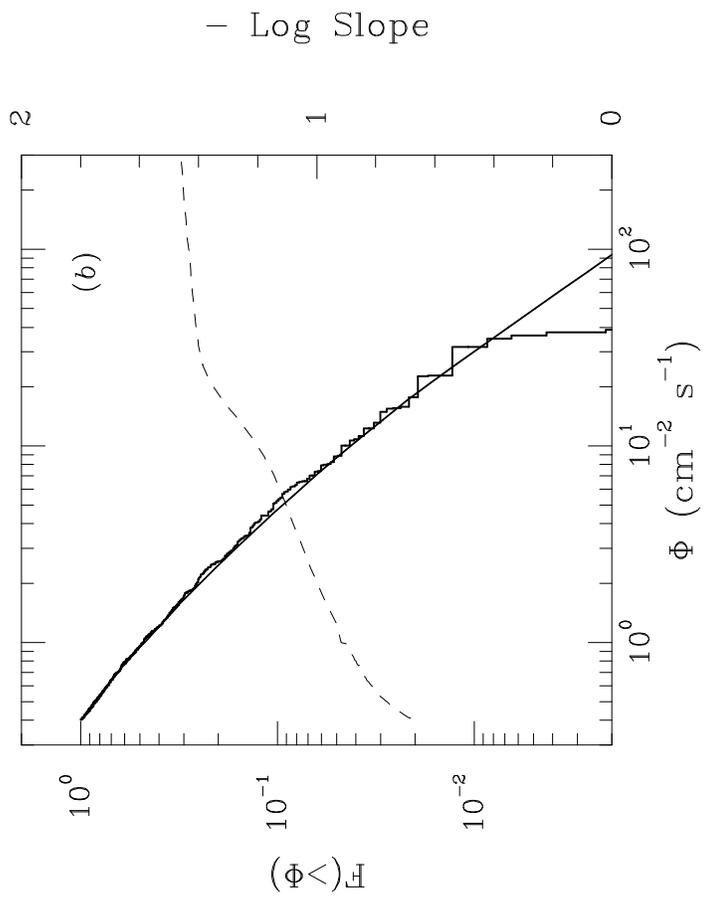
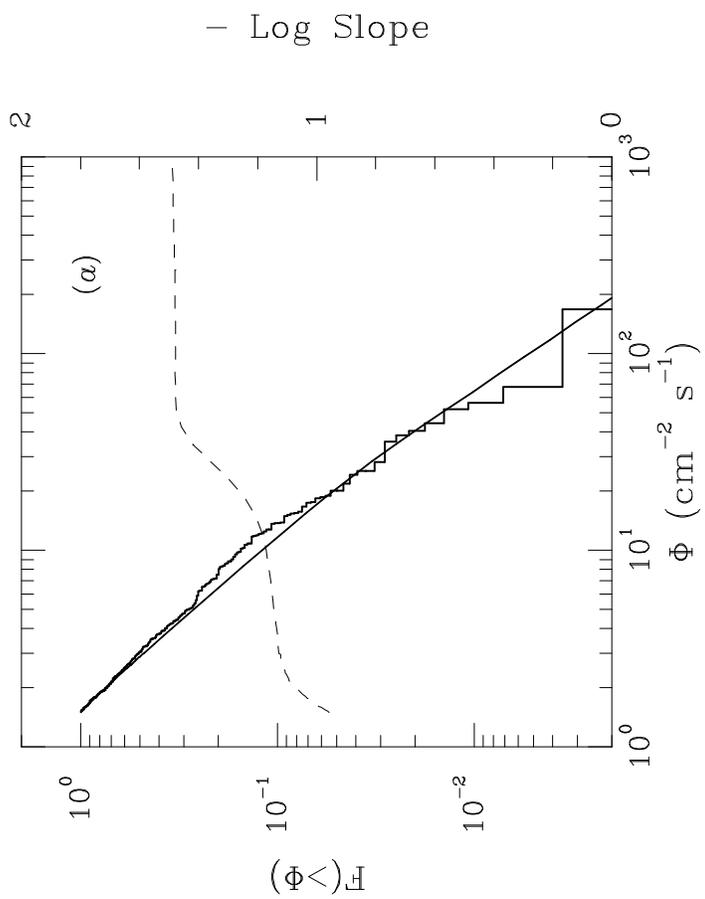
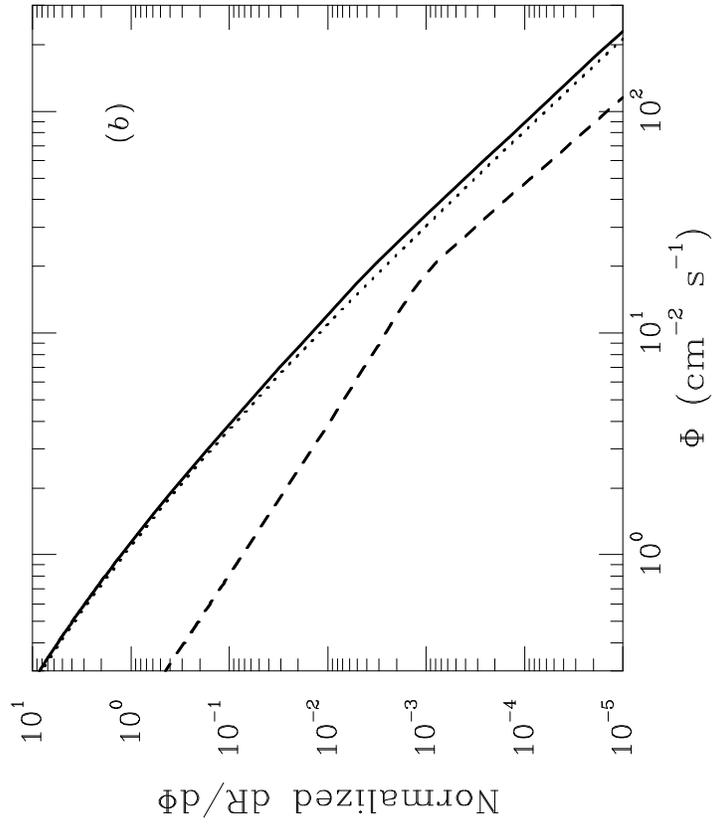
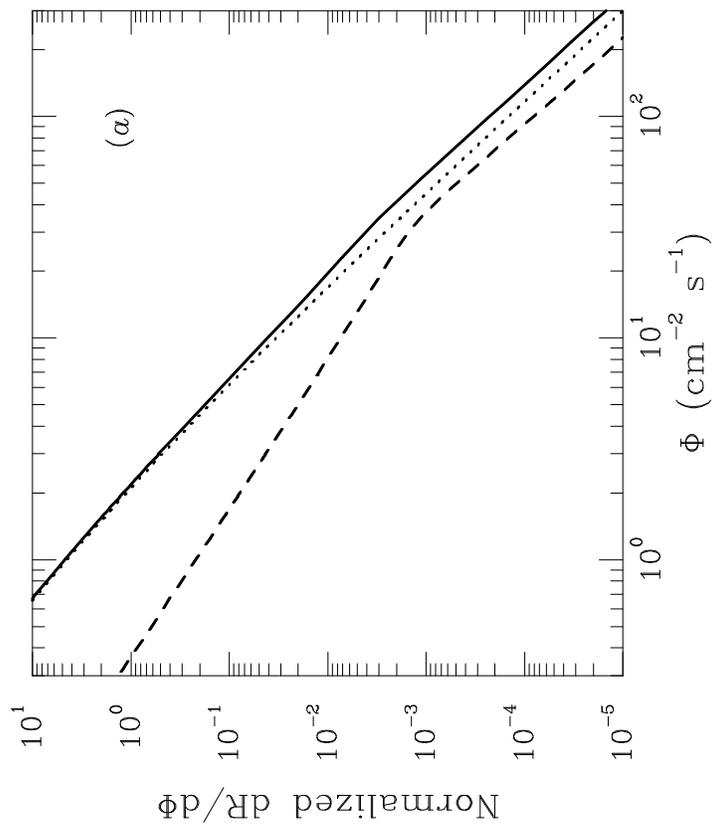
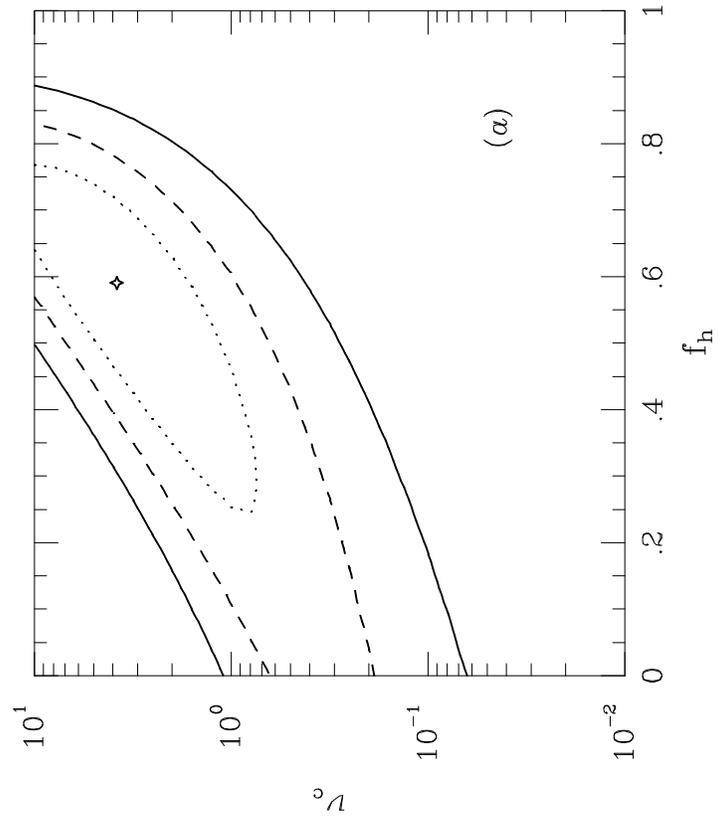
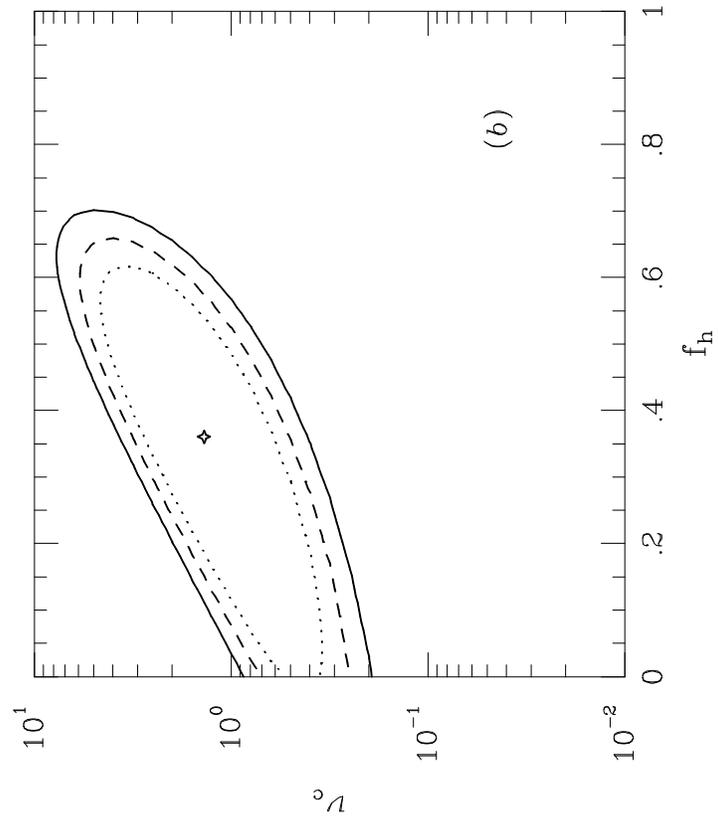
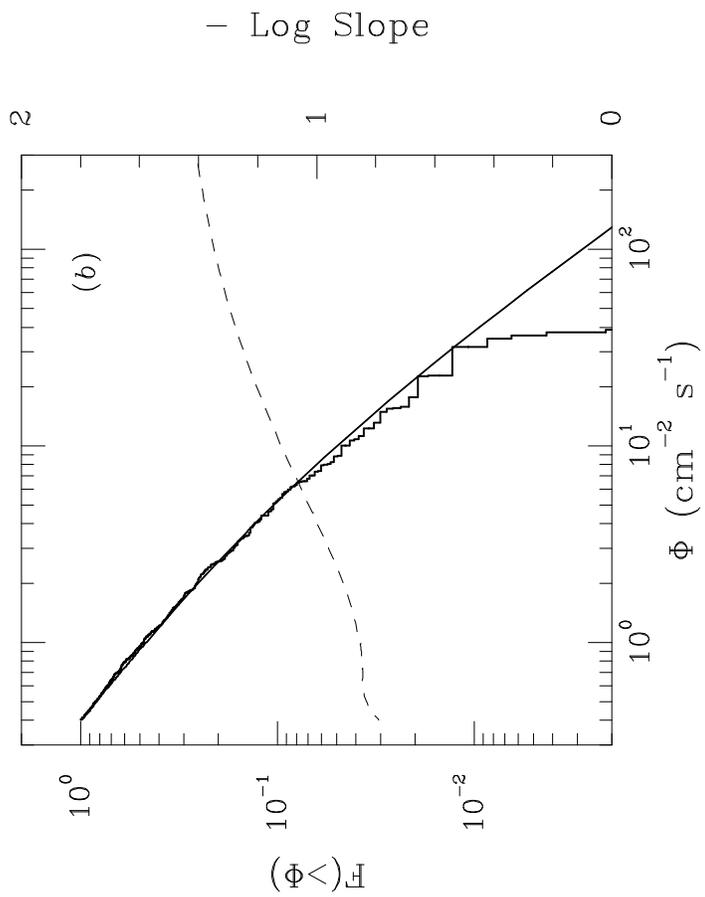
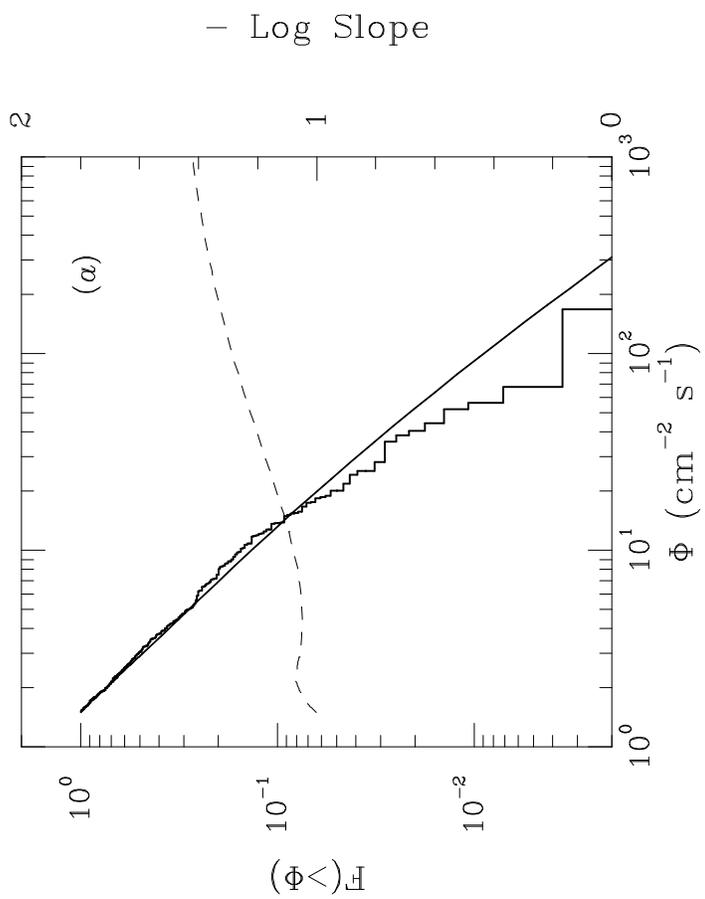
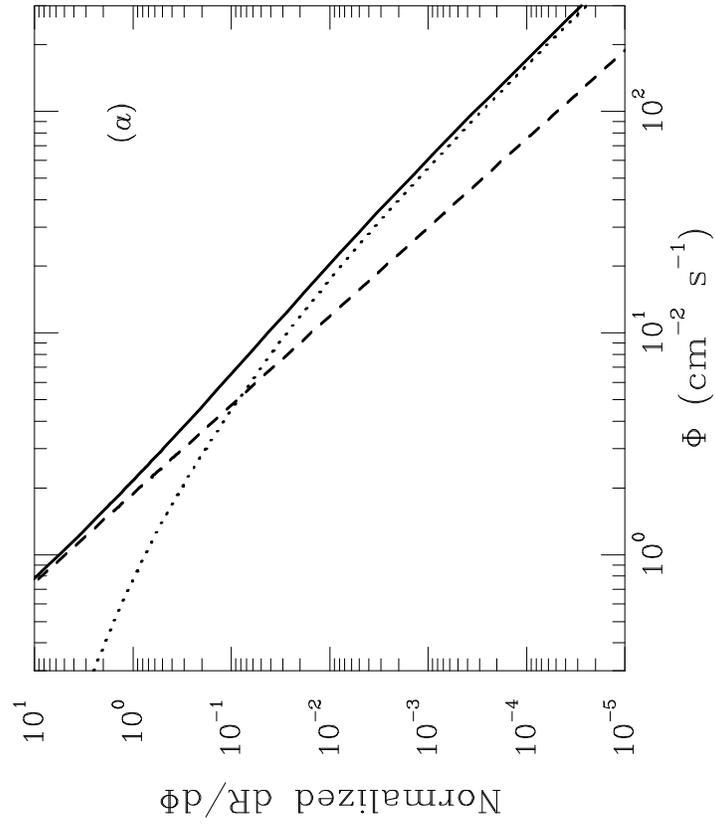
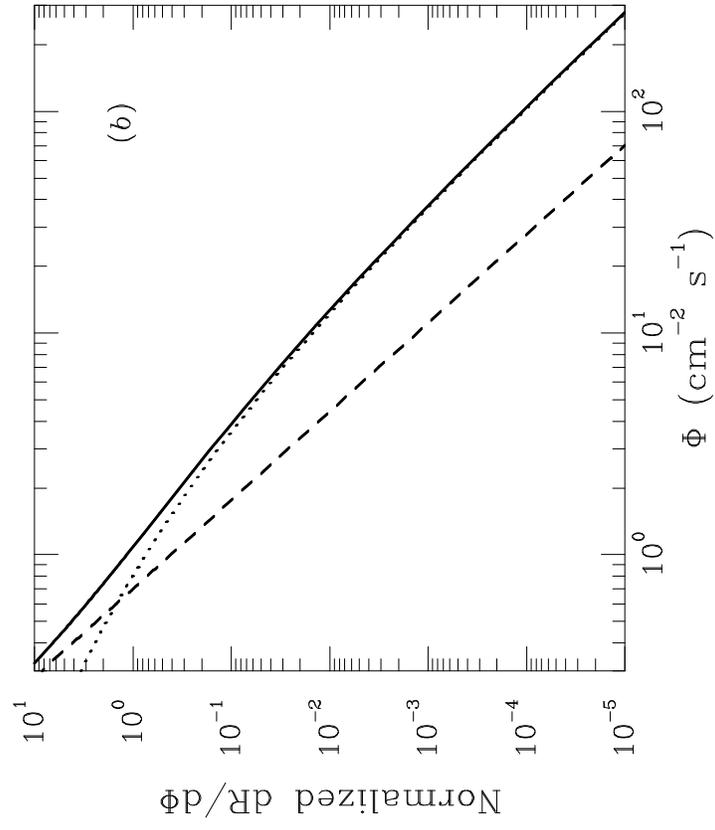
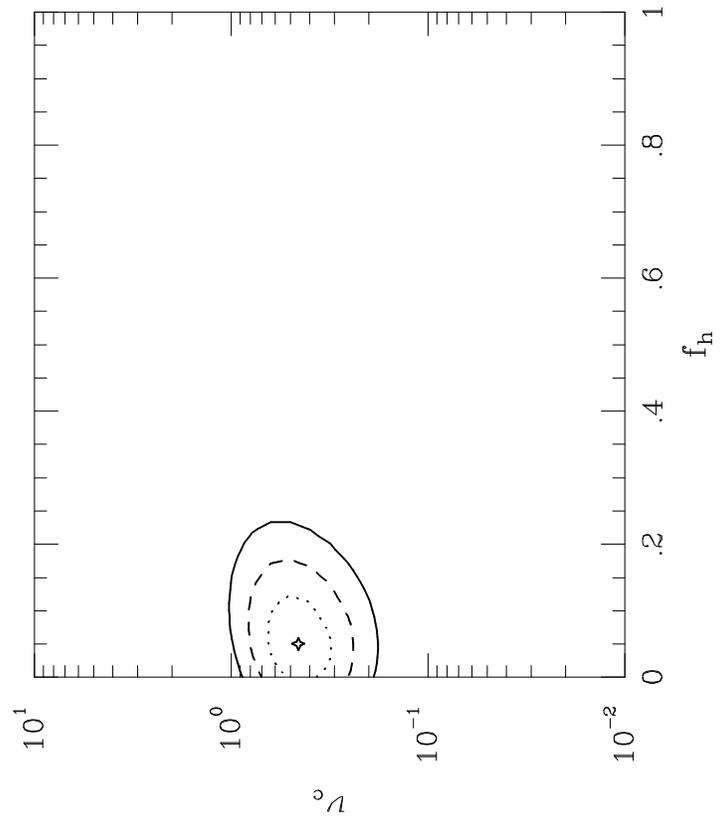
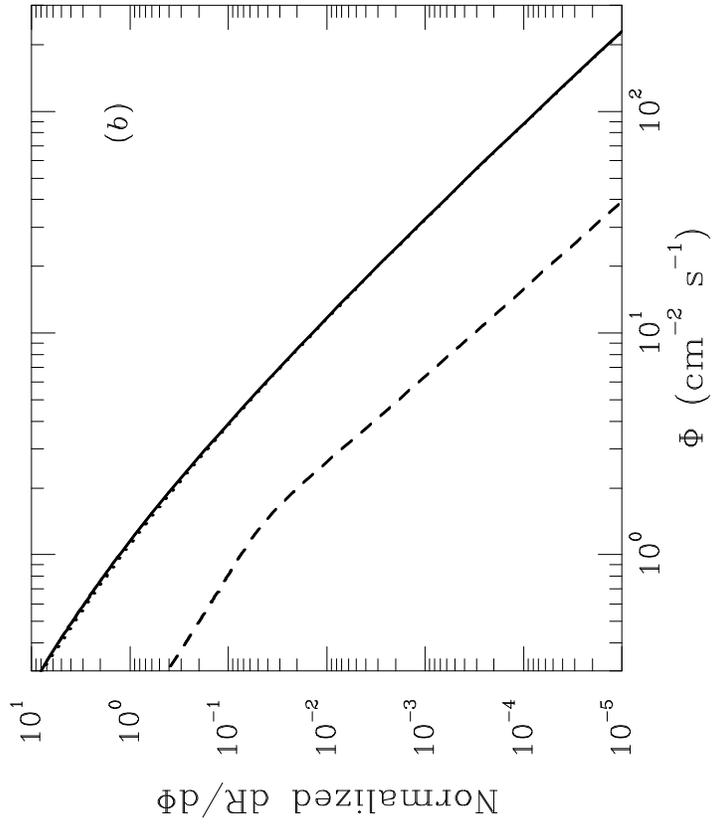
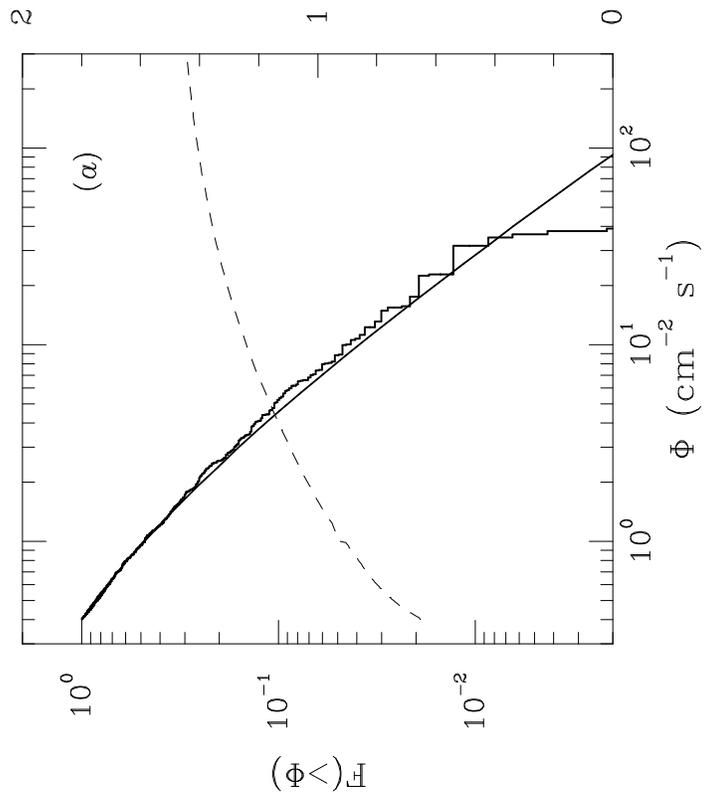
\end{document}